\documentclass[a4paper]{article}

\usepackage[english]{babel}
\usepackage[utf8x]{inputenc}
\usepackage[T1]{fontenc}

\usepackage[a4paper,top=3cm,bottom=2cm,left=3cm,right=3cm,marginparwidth=1.75cm]{geometry}

\usepackage{graphicx}
\usepackage[table]{xcolor}
\usepackage{multirow}
\usepackage{tabu}
\definecolor{darkgray}{gray}{0.5}
\definecolor{darkblue}{rgb}{0,0,0.5}
\definecolor{darkred}{rgb}{0.5,0,0}
\usepackage{tikz}
\usepackage{amsmath}
\usepackage{hyperref}
\usetikzlibrary{arrows,shapes.geometric,topaths,calc}

\begin{document}

\title{\bf Higher order molecular organisation as a source of biological function}

\author{Thomas Gaudelet, Noel Malod-Dognin and Natasa Przulj\footnote{natasa@cs.ucl.ac.uk}}
\date{University College London, Department of Computer Science, London, WC1E 6BT, United-Kingdom}

\maketitle

\begin{abstract}
\noindent\textbf{Motivation:} Molecular interactions have widely been modelled as networks. The local wiring patterns around molecules in molecular networks are linked with their biological functions. However, networks model only pairwise interactions between molecules and cannot explicitly and directly capture the higher order molecular organisation, such as protein complexes and pathways.  Hence, we ask if \textit{hypergraphs} (\textit{hypernetworks}), that directly capture entire complexes and pathways along with protein-protein interactions (PPIs), carry additional functional information beyond what can be uncovered from networks of pairwise molecular interactions.  The mathematical formalism of a hypergraph has long been known, but not often used in studying molecular networks due to the lack of sophisticated algorithms for mining the underlying biological information hidden in the wiring patterns of molecular systems modelled as hypernetworks.  \\
\textbf{Results:} 
We propose a new,\textit{ multi-scale}, protein interaction \textit{hypernetwork model} that utilizes hypergraphs to capture different scales of protein organization, including PPIs, protein complexes and pathways. In analogy to graphlets, we introduce \textit{hypergraphlets}, small, connected, non-isomorphic, induced sub-hypergraphs of a hypergraph, to quantify the local wiring patterns of these multi-scale molecular hypergraphs and to mine them for new biological information. We apply them to model the multi-scale protein networks of baker’s yeast and human and show that the higher order molecular organisation captured by these hypergraphs is strongly related to the underlying biology. Importantly, we demonstrate that our new models and data mining tools reveal different, but complementary biological information compared to classical PPI networks.  We apply our hypergraphlets to successfully predict biological functions of uncharacterised proteins.

\textbf{This article has been accepted for publication in Bioinformatics Published by Oxford University Press, \url{https://academic.oup.com/bioinformatics/article/34/17/i944/5093209}.}
\end{abstract}

\section{Introduction}

Deciphering the complex patterns of interactions between macromolecules in a cell is of crucial importance. Graph theory offers mathematical abstractions to represent and study molecular interactions. Simple \textit{graphs} (also called \textit{networks}) have been widely used to model the interactions between pairs of molecules. For instance, in Protein-Protein Interaction (PPI) networks, each node represents a protein and each edge connects a pair of proteins that can physically interact \cite{uetz2000comprehensive,ito2001comprehensive,stelzl2005human,rolland2014proteome}. Exact comparison of networks is a hard problem due to the NP-completeness of the underlying subgraph isomorphism problem \cite{cook1971complexity}. Thus, simple heuristics have been used to study PPI and other molecular networks, such as degree distribution and centralities \cite{mason2007graph}. \textit{Graphlets} quantify the local topology of a network. They are small, non-isomorphic, induced subgraphs of a larger network, which precisely characterise the local wiring patterns around each node \cite{Przulj2004,Przulj2007}. Graphlets and their statistics have since been used to compare biological networks \cite{Yaveroglu2014}, to uncover their functional organisation \cite{Przulj2004,Przulj2007,Milenkovic2008,Yaveroglu2014}, to guide network alignment algorithms \cite{kuchaiev2010topological,Malod-Dognin2015}, or to relate the wiring patterns of genes in these networks with their biological functions \cite{Milenkovic2008,Yaveroglu2014,davis2015topology}.

However, in biological systems, molecules do not interact solely in a pairwise fashion. Hence, simple graphs do not capture the multi-scale organisation of these systems \cite{Lacroix2008,Klamt2009}. In the example in Figure \ref{proteins}, we observe that the simple graph representation of the system on the left blurs the higher-order organisation. Given only the network representation on the right, one might, for instance, falsely assume that the nodes b, c, and d form a complex of three elements, while it is true that b and d form a complex, b and c form a complex, and c, d and e form a complex. 



\tikzstyle{state}=[circle,fill=black!25,minimum size=15pt,inner sep=0pt,text=black]
\tikzstyle{edge} = [draw,thick,-]
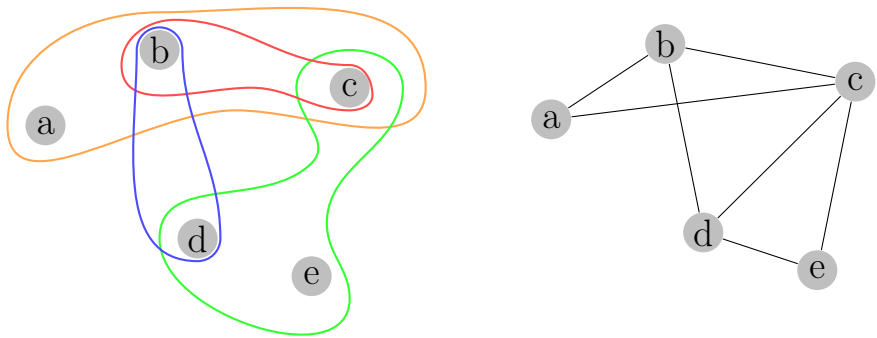
\begin{figure}[ht]
\centering
\begin{tabular}{c  c  c }
\scalebox{1}{ \begin{tikzpicture}
    \node[state] (a) at (0,2) {\Large a};
    \node[state] (b) at (1.5,3) {\Large b};
    \node[state] (c) at (4,2.5) {\Large c};
    \node[state] (d) at (2,0.5) {\Large d};
    \node[state] (e) at (3.5,0) {\Large e};

    \begin{scope}[thick,fill opacity = 0]
    \filldraw[draw=orange!70,thick] ($(a)+(-0.5,0)$) 
        to[out=90,in=180] ($(b) + (0,0.5)$) 
        to[out=0,in=90] ($(c) + (1,0)$)
        to[out=270,in=0] ($(b) + (1,-0.8)$)
        to[out=180,in=270] ($(a)+(-0.5,0)$);
    \filldraw[draw=green!80,thick] ($(d)+(-0.5,0)$)
        to[out=90,in=225] ($(c)+(-0.5,-1)$)
        to[out=45,in=270] ($(c)+(-0.7,0)$)
        to[out=90,in=180] ($(c)+(0,0.5)$)
        to[out=0,in=90] ($(c)+(0.7,0)$)
        to[out=270,in=90] ($(c)+(-0.3,-1.8)$)
        to[out=270,in=90] ($(e)+(0.5,-0.3)$)
        to[out=270,in=270] ($(d)+(-0.5,0)$);
    \filldraw[draw=red!70,thick] ($(b)+(-0.5,-0.2)$) 
        to[out=90,in=180] ($(b) + (0.2,0.4)$) 
        to[out=0,in=180] ($(c) + (0,0.3)$)
        to[out=0,in=90] ($(c) + (0.3,-0.1)$)
        to[out=270,in=0] ($(c) + (0,-0.3)$)
        to[out=180,in=0] ($(c) + (-1.3,0)$)
        to[out=180,in=270] ($(b)+(-0.5,-0.2)$);
    \filldraw[draw=blue!70,thick] ($(b)+(-0.3,0.)$) 
        to[out=90,in=180] ($(b) + (0,0.3)$) 
        to[out=0,in=90] ($(b) + (0.3,0.0)$) 
        to[out=270,in=90] ($(d) + (0.3,0)$)
        to[out=270,in=0] ($(d) + (0,-0.3)$)
        to[out=270,in=0] ($(d) + (0,-0.3)$)
        to[out=180,in=270] ($(b)+(-0.3,0.)$);
    \end{scope}
\end{tikzpicture}}

 & \hspace{1.5em} &
  
 \scalebox{1}{\begin{tikzpicture}
    \node[state] (a) at (0,2) {\Large a};
    \node[state] (b) at (1.5,3) {\Large b};
    \node[state] (c) at (4,2.5) {\Large c};
    \node[state] (d) at (2,0.5) {\Large d};
    \node[state] (e) at (3.5,0) {\Large e};
    \node (g) at (0,-1.3) {};
 
    \path (a) edge      node {} (b)
              edge      node {} (c)
          (b) edge      node {} (c)
          	  edge      node {} (d)
          (c) edge      node {} (d)
          	  edge      node {} (e)
          (d) edge      node {} (e);
          
 \end{tikzpicture}}
\end{tabular} 
\caption{\label{proteins} Illustration of a system with higher order interactions (left) and its simple graph representation (right).}
\end{figure}

A solution to overcome this limitation is to model a molecular system using hypergraphs. A \textit{hypergraph} is defined by a set of nodes, $V$, and a set of edges, $E$, called  \textit{hyperedges}, where each hyperedge corresponds to a set of interacting nodes of any size \cite{Berge1973}. This means that a simple graph is a special case of a hypergraph in which all hyperedges are sets of two nodes. The representation of the system in Figure \ref{proteins} (left) is a hypergraph. To analyse data modelled as hypergraphs, it is necessary to develop methods to mine the structure of hypergraphs. A number of simple measures from graph theory have already been extended to hypergraphs, e.g., the clustering coefficient \cite{Estrada2006}, degree distribution \cite{Latapy2008}, and centralities \cite{Estrada2006,Pearcy2014}. Approaches such as percolation and random walks \cite{Bellaachia2013,Pearcy2016} have also been extended to study hypergraphs. Hypergraphs have also been used for learning tasks, such as clustering and nodes classification \cite{Tian2009,Pelillo2013}. However, hypergraphs lack more advanced descriptors of local topology. Hence, we introduce hypergraphlets, an extension of graphlets to hypernetworks.

We investigate biological hypernetworks in which nodes are proteins and hyperedges capture PPIs, protein complexes, or signaling pathways. The main aim is to check if the topology of these hypernetwork representations of the data carries biological information that goes beyond the information that can be obtained from PPI networks. We use hypergraphlets in this investigation. 

\section{Contributions} 

We motivate studying the higher order molecular interactions as models that capture additional and different biological information than the widely studied PPI networks. We introduce hypergraphlets as a new tool that unveils the pioneering observation of the close link between the multi-scale molecular organisation and biological function and that can serve as an underlying methodology for many new tools that will be developed to further study the multi-scale organisation of molecular systems.

We analyse the hypergraph representation of protein interactions of yeast \textit{saccharomyces cerevisiae} and human and show that proteins that are similarly wired in a hypernetwork, independently of their location in the hypernetwork, tend to have similar biological functions. Also, we use the Canonical Correlation Analysis (CCA) \cite{Hardoon2004} to correlate hypergraphlets around proteins in these networks with their biological functions. The results confirm the link between the local wiring patterns of the multi-scale molecular organisation of the cell and biological functions. We use these findings to predict biological functions of uncharacterised proteins from the wiring patterns of the multi-scale molecular organisation. We validate our predictions in the literature.

\section{Materials \& Methods}

\subsection{Data \label{material}}
We consider six different networks across two species, human and baker's yeast. For each species, we consider the protein-protein interaction (PPI) network and  two hypernetworks corresponding to protein complexes and biological pathways. In all networks, nodes correspond to proteins. In a PPI network, an edge between two proteins represents a physical interaction. Depending on the hypernetwork considered, a hyperedge represents either a protein complex or a biological pathway. These data are used jointly to build hypernetworks capturing multi-scale organisation of proteins in a cell, as detailed in Section \ref{stuff} below.

The PPI data is obtained from the BioGRID database \cite{biogrid} (version 3.4.145). Both pathways hypernetworks come from the Reactome database \cite{reactome1} (accessed in April 2017). The human protein complexes are downloaded from the CORUM database \cite{Ruepp2008,Ruepp2010} (in May 2017), while the yeast protein complexes are collected from the CYC2008 database \cite{Pu2009} (last updated in 2009). Table \ref{tab:databases} gives an overview of the sizes of the data sets. 

 \begin{table}[ht]
  \centering
  
  \begin{tabular}{ c  c  c  c }
   		& Database & \# proteins & \# (hyper-) interactions  \\
        \hline
   &CORUM & 3,145 & 2,138 \\ 
    Human & Reactome & 9,466 & 1,461  \\  
   & PPI & 16,008 & 216,865\\
   \hline
   & Reactome & 1,465 & 400  \\  
   Yeast & Cyc2008 & 1,607 & 406 \\
    & PPI & 5,931 & 87,225
  \end{tabular}  
\caption{\label{tab:databases} Sizes of the data.}
\end{table}

To investigate the links between networks and biological functions, we collect gene annotations from the Gene Ontology Consortium (GO) database \cite{Consortium2015} (downloaded at the end of January 2017). For each protein, we keep only the most specific annotations that are experimentally derived. We separate the annotations based on the three categories: Biological Process (BP), Molecular Function (MF), and Cellular Component (CC).

\subsection{Hypergraphlets: the local topology of hypergraphs}

We define \textit{hypergraphlets} as small, connected, non-isomorphic, induced sub-hypergraphs of larger hypergraphs. \cite{Berge1973} defines an induced sub-hypergraph of a hypergraph $H = (V,E)$ on a set of nodes $A \subset V$ as the hypergraph $H_A$ with set of nodes $A$ and set of unique hyperedges
\begin{align}
E_{H_A} = \{e\cap A | e\in E,e\cap A \neq \emptyset\}.
\end{align}
Note that with this definition, 1-hyperedges containing only one node exist for each node. With this definition, an induced hypergraph is simple, i.e. it has no duplicated edges.

\begin{figure}[ht]
\centering
\includegraphics[width=\linewidth]{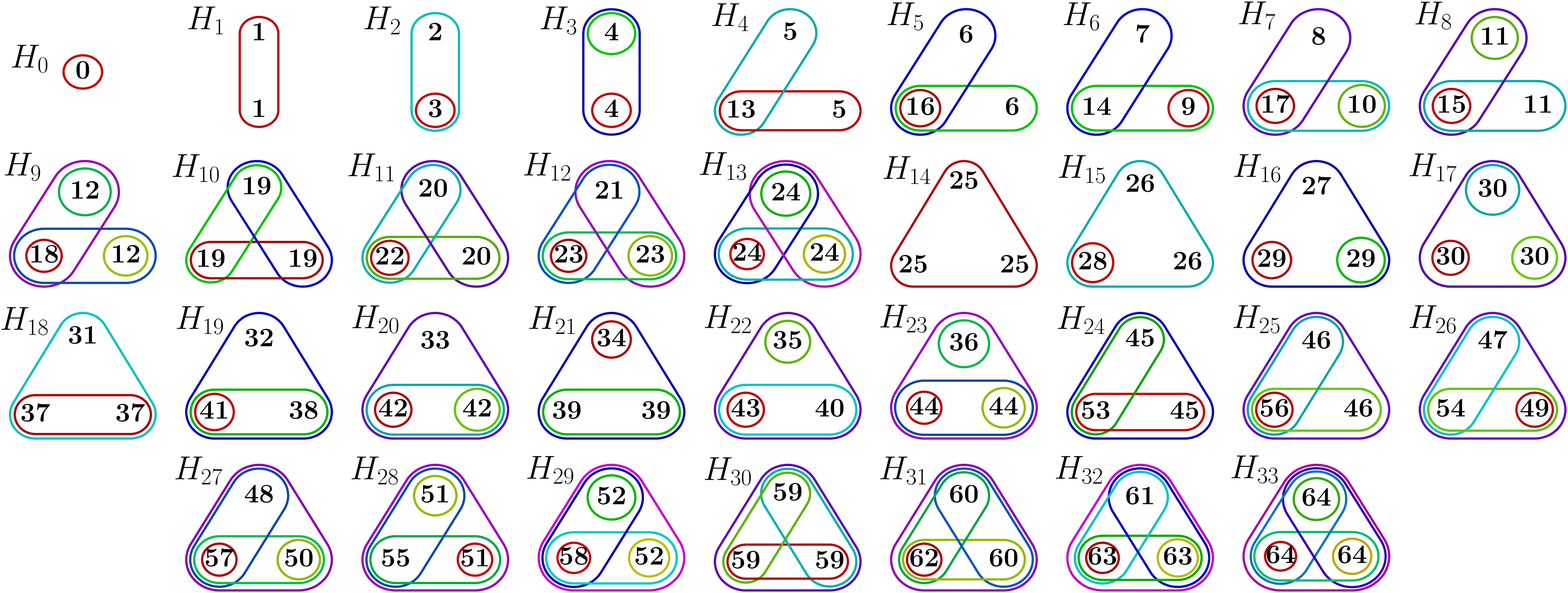}
\caption{\label{fig1} Illustration of all $1$- to $3$-node hypergraphlets ($H_{0}$ to $H_{33}$) and the $65$ orbits. Each closed set corresponds to a hyperedge and each node is represented by an integer between $0$ and $64$ corresponding to the orbit it belongs to.}
\end{figure}

Within a given hypergraph, automorphic nodes are nodes whose labels can be exchanged without changing adjacency relationships. Formally these nodes can be mapped to each other by an \textit{automorphism}, which is an isomorphism of a hypergraph with itself. An \textit{isomorphism} is a mapping of nodes of the hypergraph that preserves the adjacency of the nodes \cite{bondy1976graph}. A set of automorphic nodes form what is called an \textit{orbit}. Here, we consider all $1$- to $4$-node hypergraphlets, which contain a total of $6,369$ different orbits. For $5$-node hypergraphlets, we estimate that there are more than a hundred thousands orbits, hence we restrict ourselves to $4$-node hypergraphlets. In Figure \ref{fig1}, we illustrate all $65$ orbits that occur in the $1$- to $3$-node hypergraphlets. 

Analogous to graphlets, we use hypergraphlet orbits to quantify the wiring patterns around each node in a hypergraph. For each orbit $i$ in hypergraphlet $h$, we define the $i^{th}$\textit{ hypergraphlet degree} of a node in the hypergraph $H$ as the number of hypergraphlet orbits $i$ that the node touches. 

For each node in a hypergraph, we compute all $6,369$ hypergraphlet degrees resulting in a $6,369$-dimensional vector where entry $i$ corresponds to the $i^{th}$ hypergraphlet degree of the node. We term this vector capturing the local wiring around a node the \textit{Hypergraphlet Degree Vector (HDV)}.

Considering a hypergraph with $n$ nodes, with maximal hyperedge of size $l$ and with maximal degree of a node $d$, where the \textit{degree} of a node corresponds to the number of hyperedges that contain it, an upper bound on the complexity of counting all $1$- to $k$-node hypergraphlets is $O(n(ld)^{k-1})$.

\cite{Lugo-Martinez2014} introduced an alternative definition of hypergraphlets in the context of binary classification problems. They define kernels based on their definition of hypergraphlets and use support vector machines to classify the proteins. The key difference with our definition of hypergraphlets is that they do not consider the hypergraphlets of a hypergraph as \textit{induced} sub-hypergraphs, thus ignoring some overlaps between hyperedges \cite{Lugo-Martinez2014}. In particular, in the first step, they ignore all hyperedges containing more than four nodes. Instead, hyperedges with more than four nodes are taken into consideration independently in the second step, which decomposes a hyperedge of size $n>4$ into the $\binom{n}{4}$ subsets of four nodes. Hence, with their definition and counting process, an important part of the topology of the hypernetwork is overlooked and therefore topological information is lost, which motivates our redefinition that is also a direct extension of the definition of graphlets for simple graphs. However, we could not compare the two approaches, as their implementation is not publicly available and they recently agreed with us that their definition needed to be changed to alleviate these issues\footnote{Personal communication.}.

\subsection{Topological distance\label{dist}}

We define a distance measure to compare the wiring patterns of two vertices in a hypernetwork (or network, depending on the model considered) as follows. Consider a set of proteins $P = \{p_1,p_2,\ldots,p_m\}$ and let $M$ be the matrix representing our data where row $i$ corresponds to the HDV (or GDV) of  protein $p_i$. Then, we define the distance, $\delta$, between two proteins $p_i$ and $p_j$ as
\begin{align}\label{eq}
	\delta(p_i,p_j) = \left(\sum_{k \in K} \frac{\left(\log\left(M_{ik}\right) - \log\left(M_{jk}\right)\right)^2}{\sigma_k}\right)^{\frac{1}{2}},
\end{align}
where $K$ corresponds to the set of orbits considered, $M_{ik}$ denotes the entry of $M$ on the $i^{th}$ row and $k^{th}$ column, and $\sigma_k$ denotes the standard deviation of the distribution of the $k^{th}$ hypergraphlet (or graphlet) orbit degree across our set of data value. Note that to reduce the impact of very large orbit counts we apply to $M$ an element-wise log transformation.

\subsection{Linking local structure to function}

We explore two ways to evaluate the link between the local structure of a molecular network and the biological functions of its molecules. First, we cluster the nodes based on the similarity of their wiring patterns defined in Section \ref{dist}, and we do the enrichment analysis of the resulting clusters (Section \ref{EnrSec}). Second, we use CCA to test if biological functions tend to be characterised by specific wiring patterns (Section \ref{CCASec}).

\subsubsection{Cluster enrichment\label{EnrSec}}
We cluster proteins that are similarly wired in a graph or a hypergraph as measured by distance $\delta$ (see Equation \ref{eq}) and test if the proteins within the same cluster share GO functions.

Clusters are obtained by using k-means method \cite{hartigan1979algorithm} based on the distance defined in Equation \ref{eq}. For each of various numbers of clusters, $k$, we run the clustering algorithm $20$ times to account for the randomness in the k-means algorithm. For each clustering, we compute the enrichment of clusters in biological annotations for each GO category with correction for multiple hypothesis testing \cite{benjamini1995controlling}. We consider a cluster enriched if at least one GO annotation is significantly enriched in the cluster (p-value $< 5\%$). For each value of $k$, we also compute the average of Sum of Squared Error (SSE) and the Normalised Mutual Information (NMI) \cite{Vinh2010} considering all $20$ repeats. SSE gives a measure of how close proteins within a cluster are on average according to our similarity measure, while NMI evaluates the stability of the clustering across the $20$ runs, i.e. if proteins are consistently clustered together or apart. Then, we use `` the elbow'' analysis of the SSE and NMI with respect to $k$ to choose the optimal number of clusters. For the resulting number of clusters, we select the clustering giving the highest percentage enrichment across the $20$ runs of k-means for each GO category. We test the significance of the enrichment with random permutation tests: we keep the same number and size of clusters and randomly assign proteins to each cluster and measure the enrichments of the resulting clusters. We repeat this process $1,000$ times and compute the significance.

To see whether the two models, networks and hypernetworks, harbour the  same or different but complementary biological information, at least  to the extent that it can be uncovered by the proposed methodologies, we measure Adjusted Mutual Information (AMI) \cite{Vinh2010} of the clusters and Jaccard Index \cite{jaccard1912distribution} of the enriched annotations in the clusters. AMI is a variation of Mutual Information (MI) used to compare two clusterings. It measures if any pair of proteins is consistently clustered together or apart in both clusterings adjusting for chance. The Jaccard Index gives a measure of the overlap between the two sets of GO annotations.

\subsubsection{Canonical correlation analysis\label{CCASec}}

CCA is used to infer correlations between two sets of features, $X$ and $Y$. Consider features $X = \left(X_1, ..., X_n\right)$ and $Y = \left(Y_1, ..., Y_m\right)$ over the same elements. Then CCA will identify $K$ pairs $(\mathcal{L}_X^k,\mathcal{L}_Y^k)$, called \textit{canonical variates},  of linear combinations of features of $X$ and of features of $Y$, with $K = \min(m,n)$, such that the correlations of $\mathcal{L}_x^k$ and $\mathcal{L}_y^k$ are maximal over all $k$. Each canonical variate is associated a score corresponding to the correlation between its two linear combinations.

In our case, the elements are proteins, the first set of features corresponds to the wiring patterns of proteins in networks or hypernetworks, and the second to the biological functions of proteins from GO. As mentioned above, each protein (node) has a GDV from the PPI network and an  HDV from the hypernetwork. Hence, we have two matrices of topological features where entries $(i,j)$ correspond to the $j^{th}$ orbit degree of protein $i$. Also, we associate to each protein three vectors of GO annotations, one for each of the categories: BP, MF, and CC. In each of these vectors, an entry is equal to $1$ if the gene is annotated with the corresponding GO term, and $0$ otherwise. Hence, we form three matrices of biological features, where entries $(i,j)$ correspond to the presence or absence of GO annotation $j$ for protein $i$.

We compute CCA for each combination of topological features and biological annotations to uncover topology-function relationships in the data.

\subsection{Summary of the analysis \label{stuff}}

As stated above, our main aim is to examine if modeling the higher order of molecular organisation harbours additional biological information and to demonstrate that the wiring patterns of biological hypernetworks are strongly linked to the underlying biology.\\

\begin{figure}[ht]
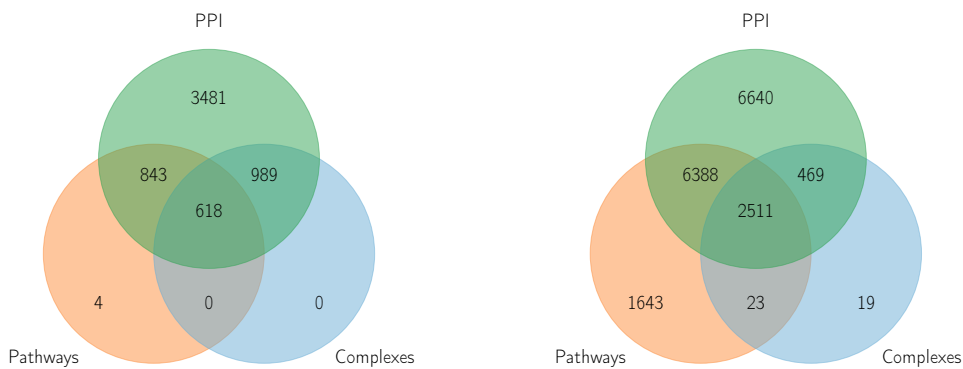

\centering
\begin{tabular}{c c c}
\centering
\scalebox{0.25}{\begin{pgfpicture}%
\pgfpathrectangle{\pgfpointorigin}{\pgfqpoint{10.000000in}{9.580000in}}%
\pgfusepath{use as bounding box, clip}%
\begin{pgfscope}%
\pgfsetbuttcap%
\pgfsetmiterjoin%
\definecolor{currentfill}{rgb}{1.000000,1.000000,1.000000}%
\pgfsetfillcolor{currentfill}%
\pgfsetlinewidth{0.000000pt}%
\definecolor{currentstroke}{rgb}{1.000000,1.000000,1.000000}%
\pgfsetstrokecolor{currentstroke}%
\pgfsetdash{}{0pt}%
\pgfpathmoveto{\pgfqpoint{0.000000in}{0.000000in}}%
\pgfpathlineto{\pgfqpoint{10.000000in}{0.000000in}}%
\pgfpathlineto{\pgfqpoint{10.000000in}{9.580000in}}%
\pgfpathlineto{\pgfqpoint{0.000000in}{9.580000in}}%
\pgfpathclose%
\pgfusepath{fill}%
\end{pgfscope}%
\begin{pgfscope}%
\pgfpathrectangle{\pgfqpoint{0.425000in}{0.215000in}}{\pgfqpoint{9.150000in}{9.150000in}} %
\pgfusepath{clip}%
\pgfsetbuttcap%
\pgfsetmiterjoin%
\definecolor{currentfill}{rgb}{0.992157,0.552941,0.235294}%
\pgfsetfillcolor{currentfill}%
\pgfsetfillopacity{0.500000}%
\pgfsetlinewidth{1.003750pt}%
\definecolor{currentstroke}{rgb}{0.992157,0.552941,0.235294}%
\pgfsetstrokecolor{currentstroke}%
\pgfsetstrokeopacity{0.500000}%
\pgfsetdash{}{0pt}%
\pgfpathmoveto{\pgfqpoint{3.856250in}{1.358750in}}%
\pgfpathcurveto{\pgfqpoint{4.462902in}{1.358750in}}{\pgfqpoint{5.044789in}{1.599775in}}{\pgfqpoint{5.473757in}{2.028743in}}%
\pgfpathcurveto{\pgfqpoint{5.902725in}{2.457711in}}{\pgfqpoint{6.143750in}{3.039598in}}{\pgfqpoint{6.143750in}{3.646250in}}%
\pgfpathcurveto{\pgfqpoint{6.143750in}{4.252902in}}{\pgfqpoint{5.902725in}{4.834789in}}{\pgfqpoint{5.473757in}{5.263757in}}%
\pgfpathcurveto{\pgfqpoint{5.044789in}{5.692725in}}{\pgfqpoint{4.462902in}{5.933750in}}{\pgfqpoint{3.856250in}{5.933750in}}%
\pgfpathcurveto{\pgfqpoint{3.249598in}{5.933750in}}{\pgfqpoint{2.667711in}{5.692725in}}{\pgfqpoint{2.238743in}{5.263757in}}%
\pgfpathcurveto{\pgfqpoint{1.809775in}{4.834789in}}{\pgfqpoint{1.568750in}{4.252902in}}{\pgfqpoint{1.568750in}{3.646250in}}%
\pgfpathcurveto{\pgfqpoint{1.568750in}{3.039598in}}{\pgfqpoint{1.809775in}{2.457711in}}{\pgfqpoint{2.238743in}{2.028743in}}%
\pgfpathcurveto{\pgfqpoint{2.667711in}{1.599775in}}{\pgfqpoint{3.249598in}{1.358750in}}{\pgfqpoint{3.856250in}{1.358750in}}%
\pgfpathclose%
\pgfusepath{stroke,fill}%
\end{pgfscope}%
\begin{pgfscope}%
\pgfpathrectangle{\pgfqpoint{0.425000in}{0.215000in}}{\pgfqpoint{9.150000in}{9.150000in}} %
\pgfusepath{clip}%
\pgfsetbuttcap%
\pgfsetmiterjoin%
\definecolor{currentfill}{rgb}{0.419608,0.682353,0.839216}%
\pgfsetfillcolor{currentfill}%
\pgfsetfillopacity{0.500000}%
\pgfsetlinewidth{1.003750pt}%
\definecolor{currentstroke}{rgb}{0.419608,0.682353,0.839216}%
\pgfsetstrokecolor{currentstroke}%
\pgfsetstrokeopacity{0.500000}%
\pgfsetdash{}{0pt}%
\pgfpathmoveto{\pgfqpoint{6.143750in}{1.358750in}}%
\pgfpathcurveto{\pgfqpoint{6.750402in}{1.358750in}}{\pgfqpoint{7.332289in}{1.599775in}}{\pgfqpoint{7.761257in}{2.028743in}}%
\pgfpathcurveto{\pgfqpoint{8.190225in}{2.457711in}}{\pgfqpoint{8.431250in}{3.039598in}}{\pgfqpoint{8.431250in}{3.646250in}}%
\pgfpathcurveto{\pgfqpoint{8.431250in}{4.252902in}}{\pgfqpoint{8.190225in}{4.834789in}}{\pgfqpoint{7.761257in}{5.263757in}}%
\pgfpathcurveto{\pgfqpoint{7.332289in}{5.692725in}}{\pgfqpoint{6.750402in}{5.933750in}}{\pgfqpoint{6.143750in}{5.933750in}}%
\pgfpathcurveto{\pgfqpoint{5.537098in}{5.933750in}}{\pgfqpoint{4.955211in}{5.692725in}}{\pgfqpoint{4.526243in}{5.263757in}}%
\pgfpathcurveto{\pgfqpoint{4.097275in}{4.834789in}}{\pgfqpoint{3.856250in}{4.252902in}}{\pgfqpoint{3.856250in}{3.646250in}}%
\pgfpathcurveto{\pgfqpoint{3.856250in}{3.039598in}}{\pgfqpoint{4.097275in}{2.457711in}}{\pgfqpoint{4.526243in}{2.028743in}}%
\pgfpathcurveto{\pgfqpoint{4.955211in}{1.599775in}}{\pgfqpoint{5.537098in}{1.358750in}}{\pgfqpoint{6.143750in}{1.358750in}}%
\pgfpathclose%
\pgfusepath{stroke,fill}%
\end{pgfscope}%
\begin{pgfscope}%
\pgfpathrectangle{\pgfqpoint{0.425000in}{0.215000in}}{\pgfqpoint{9.150000in}{9.150000in}} %
\pgfusepath{clip}%
\pgfsetbuttcap%
\pgfsetmiterjoin%
\definecolor{currentfill}{rgb}{0.192157,0.639216,0.329412}%
\pgfsetfillcolor{currentfill}%
\pgfsetfillopacity{0.500000}%
\pgfsetlinewidth{1.003750pt}%
\definecolor{currentstroke}{rgb}{0.192157,0.639216,0.329412}%
\pgfsetstrokecolor{currentstroke}%
\pgfsetstrokeopacity{0.500000}%
\pgfsetdash{}{0pt}%
\pgfpathmoveto{\pgfqpoint{5.000000in}{3.339783in}}%
\pgfpathcurveto{\pgfqpoint{5.606652in}{3.339783in}}{\pgfqpoint{6.188539in}{3.580809in}}{\pgfqpoint{6.617507in}{4.009776in}}%
\pgfpathcurveto{\pgfqpoint{7.046475in}{4.438744in}}{\pgfqpoint{7.287500in}{5.020631in}}{\pgfqpoint{7.287500in}{5.627283in}}%
\pgfpathcurveto{\pgfqpoint{7.287500in}{6.233935in}}{\pgfqpoint{7.046475in}{6.815822in}}{\pgfqpoint{6.617507in}{7.244790in}}%
\pgfpathcurveto{\pgfqpoint{6.188539in}{7.673758in}}{\pgfqpoint{5.606652in}{7.914783in}}{\pgfqpoint{5.000000in}{7.914783in}}%
\pgfpathcurveto{\pgfqpoint{4.393348in}{7.914783in}}{\pgfqpoint{3.811461in}{7.673758in}}{\pgfqpoint{3.382493in}{7.244790in}}%
\pgfpathcurveto{\pgfqpoint{2.953525in}{6.815822in}}{\pgfqpoint{2.712500in}{6.233935in}}{\pgfqpoint{2.712500in}{5.627283in}}%
\pgfpathcurveto{\pgfqpoint{2.712500in}{5.020631in}}{\pgfqpoint{2.953525in}{4.438744in}}{\pgfqpoint{3.382493in}{4.009776in}}%
\pgfpathcurveto{\pgfqpoint{3.811461in}{3.580809in}}{\pgfqpoint{4.393348in}{3.339783in}}{\pgfqpoint{5.000000in}{3.339783in}}%
\pgfpathclose%
\pgfusepath{stroke,fill}%
\end{pgfscope}%
\begin{pgfscope}%
\pgftext[x=2.712500in,y=2.502500in,,base]{\sffamily\fontsize{30.000000}{36.000000}\selectfont 4}%
\end{pgfscope}%
\begin{pgfscope}%
\pgftext[x=7.287500in,y=2.502500in,,base]{\sffamily\fontsize{30.000000}{36.000000}\selectfont 0}%
\end{pgfscope}%
\begin{pgfscope}%
\pgftext[x=5.000000in,y=6.771033in,,base]{\sffamily\fontsize{30.000000}{36.000000}\selectfont 3481}%
\end{pgfscope}%
\begin{pgfscope}%
\pgftext[x=5.000000in,y=2.502500in,,base]{\sffamily\fontsize{30.000000}{36.000000}\selectfont 0}%
\end{pgfscope}%
\begin{pgfscope}%
\pgftext[x=3.856250in,y=5.171250in,,base]{\sffamily\fontsize{30.000000}{36.000000}\selectfont 843}%
\end{pgfscope}%
\begin{pgfscope}%
\pgftext[x=6.143750in,y=5.171250in,,base]{\sffamily\fontsize{30.000000}{36.000000}\selectfont 989}%
\end{pgfscope}%
\begin{pgfscope}%
\pgftext[x=5.000000in,y=4.408750in,,base]{\sffamily\fontsize{30.000000}{36.000000}\selectfont 618}%
\end{pgfscope}%
\begin{pgfscope}%
\pgftext[x=1.568750in,y=1.358750in,,base]{\sffamily\fontsize{30.000000}{36.000000}\selectfont Pathways}%
\end{pgfscope}%
\begin{pgfscope}%
\pgftext[x=8.431250in,y=1.358750in,,base]{\sffamily\fontsize{30.000000}{36.000000}\selectfont Complexes}%
\end{pgfscope}%
\begin{pgfscope}%
\pgftext[x=5.000000in,y=8.372283in,,base]{\sffamily\fontsize{30.000000}{36.000000}\selectfont PPI}%
\end{pgfscope}%
\end{pgfpicture}} & & \scalebox{0.25}{\begin{pgfpicture}%
\pgfpathrectangle{\pgfpointorigin}{\pgfqpoint{10.000000in}{9.580000in}}%
\pgfusepath{use as bounding box, clip}%
\begin{pgfscope}%
\pgfsetbuttcap%
\pgfsetmiterjoin%
\definecolor{currentfill}{rgb}{1.000000,1.000000,1.000000}%
\pgfsetfillcolor{currentfill}%
\pgfsetlinewidth{0.000000pt}%
\definecolor{currentstroke}{rgb}{1.000000,1.000000,1.000000}%
\pgfsetstrokecolor{currentstroke}%
\pgfsetdash{}{0pt}%
\pgfpathmoveto{\pgfqpoint{0.000000in}{0.000000in}}%
\pgfpathlineto{\pgfqpoint{10.000000in}{0.000000in}}%
\pgfpathlineto{\pgfqpoint{10.000000in}{9.580000in}}%
\pgfpathlineto{\pgfqpoint{0.000000in}{9.580000in}}%
\pgfpathclose%
\pgfusepath{fill}%
\end{pgfscope}%
\begin{pgfscope}%
\pgfpathrectangle{\pgfqpoint{0.425000in}{0.215000in}}{\pgfqpoint{9.150000in}{9.150000in}} %
\pgfusepath{clip}%
\pgfsetbuttcap%
\pgfsetmiterjoin%
\definecolor{currentfill}{rgb}{0.992157,0.552941,0.235294}%
\pgfsetfillcolor{currentfill}%
\pgfsetfillopacity{0.500000}%
\pgfsetlinewidth{1.003750pt}%
\definecolor{currentstroke}{rgb}{0.992157,0.552941,0.235294}%
\pgfsetstrokecolor{currentstroke}%
\pgfsetstrokeopacity{0.500000}%
\pgfsetdash{}{0pt}%
\pgfpathmoveto{\pgfqpoint{3.856250in}{1.358750in}}%
\pgfpathcurveto{\pgfqpoint{4.462902in}{1.358750in}}{\pgfqpoint{5.044789in}{1.599775in}}{\pgfqpoint{5.473757in}{2.028743in}}%
\pgfpathcurveto{\pgfqpoint{5.902725in}{2.457711in}}{\pgfqpoint{6.143750in}{3.039598in}}{\pgfqpoint{6.143750in}{3.646250in}}%
\pgfpathcurveto{\pgfqpoint{6.143750in}{4.252902in}}{\pgfqpoint{5.902725in}{4.834789in}}{\pgfqpoint{5.473757in}{5.263757in}}%
\pgfpathcurveto{\pgfqpoint{5.044789in}{5.692725in}}{\pgfqpoint{4.462902in}{5.933750in}}{\pgfqpoint{3.856250in}{5.933750in}}%
\pgfpathcurveto{\pgfqpoint{3.249598in}{5.933750in}}{\pgfqpoint{2.667711in}{5.692725in}}{\pgfqpoint{2.238743in}{5.263757in}}%
\pgfpathcurveto{\pgfqpoint{1.809775in}{4.834789in}}{\pgfqpoint{1.568750in}{4.252902in}}{\pgfqpoint{1.568750in}{3.646250in}}%
\pgfpathcurveto{\pgfqpoint{1.568750in}{3.039598in}}{\pgfqpoint{1.809775in}{2.457711in}}{\pgfqpoint{2.238743in}{2.028743in}}%
\pgfpathcurveto{\pgfqpoint{2.667711in}{1.599775in}}{\pgfqpoint{3.249598in}{1.358750in}}{\pgfqpoint{3.856250in}{1.358750in}}%
\pgfpathclose%
\pgfusepath{stroke,fill}%
\end{pgfscope}%
\begin{pgfscope}%
\pgfpathrectangle{\pgfqpoint{0.425000in}{0.215000in}}{\pgfqpoint{9.150000in}{9.150000in}} %
\pgfusepath{clip}%
\pgfsetbuttcap%
\pgfsetmiterjoin%
\definecolor{currentfill}{rgb}{0.419608,0.682353,0.839216}%
\pgfsetfillcolor{currentfill}%
\pgfsetfillopacity{0.500000}%
\pgfsetlinewidth{1.003750pt}%
\definecolor{currentstroke}{rgb}{0.419608,0.682353,0.839216}%
\pgfsetstrokecolor{currentstroke}%
\pgfsetstrokeopacity{0.500000}%
\pgfsetdash{}{0pt}%
\pgfpathmoveto{\pgfqpoint{6.143750in}{1.358750in}}%
\pgfpathcurveto{\pgfqpoint{6.750402in}{1.358750in}}{\pgfqpoint{7.332289in}{1.599775in}}{\pgfqpoint{7.761257in}{2.028743in}}%
\pgfpathcurveto{\pgfqpoint{8.190225in}{2.457711in}}{\pgfqpoint{8.431250in}{3.039598in}}{\pgfqpoint{8.431250in}{3.646250in}}%
\pgfpathcurveto{\pgfqpoint{8.431250in}{4.252902in}}{\pgfqpoint{8.190225in}{4.834789in}}{\pgfqpoint{7.761257in}{5.263757in}}%
\pgfpathcurveto{\pgfqpoint{7.332289in}{5.692725in}}{\pgfqpoint{6.750402in}{5.933750in}}{\pgfqpoint{6.143750in}{5.933750in}}%
\pgfpathcurveto{\pgfqpoint{5.537098in}{5.933750in}}{\pgfqpoint{4.955211in}{5.692725in}}{\pgfqpoint{4.526243in}{5.263757in}}%
\pgfpathcurveto{\pgfqpoint{4.097275in}{4.834789in}}{\pgfqpoint{3.856250in}{4.252902in}}{\pgfqpoint{3.856250in}{3.646250in}}%
\pgfpathcurveto{\pgfqpoint{3.856250in}{3.039598in}}{\pgfqpoint{4.097275in}{2.457711in}}{\pgfqpoint{4.526243in}{2.028743in}}%
\pgfpathcurveto{\pgfqpoint{4.955211in}{1.599775in}}{\pgfqpoint{5.537098in}{1.358750in}}{\pgfqpoint{6.143750in}{1.358750in}}%
\pgfpathclose%
\pgfusepath{stroke,fill}%
\end{pgfscope}%
\begin{pgfscope}%
\pgfpathrectangle{\pgfqpoint{0.425000in}{0.215000in}}{\pgfqpoint{9.150000in}{9.150000in}} %
\pgfusepath{clip}%
\pgfsetbuttcap%
\pgfsetmiterjoin%
\definecolor{currentfill}{rgb}{0.192157,0.639216,0.329412}%
\pgfsetfillcolor{currentfill}%
\pgfsetfillopacity{0.500000}%
\pgfsetlinewidth{1.003750pt}%
\definecolor{currentstroke}{rgb}{0.192157,0.639216,0.329412}%
\pgfsetstrokecolor{currentstroke}%
\pgfsetstrokeopacity{0.500000}%
\pgfsetdash{}{0pt}%
\pgfpathmoveto{\pgfqpoint{5.000000in}{3.339783in}}%
\pgfpathcurveto{\pgfqpoint{5.606652in}{3.339783in}}{\pgfqpoint{6.188539in}{3.580809in}}{\pgfqpoint{6.617507in}{4.009776in}}%
\pgfpathcurveto{\pgfqpoint{7.046475in}{4.438744in}}{\pgfqpoint{7.287500in}{5.020631in}}{\pgfqpoint{7.287500in}{5.627283in}}%
\pgfpathcurveto{\pgfqpoint{7.287500in}{6.233935in}}{\pgfqpoint{7.046475in}{6.815822in}}{\pgfqpoint{6.617507in}{7.244790in}}%
\pgfpathcurveto{\pgfqpoint{6.188539in}{7.673758in}}{\pgfqpoint{5.606652in}{7.914783in}}{\pgfqpoint{5.000000in}{7.914783in}}%
\pgfpathcurveto{\pgfqpoint{4.393348in}{7.914783in}}{\pgfqpoint{3.811461in}{7.673758in}}{\pgfqpoint{3.382493in}{7.244790in}}%
\pgfpathcurveto{\pgfqpoint{2.953525in}{6.815822in}}{\pgfqpoint{2.712500in}{6.233935in}}{\pgfqpoint{2.712500in}{5.627283in}}%
\pgfpathcurveto{\pgfqpoint{2.712500in}{5.020631in}}{\pgfqpoint{2.953525in}{4.438744in}}{\pgfqpoint{3.382493in}{4.009776in}}%
\pgfpathcurveto{\pgfqpoint{3.811461in}{3.580809in}}{\pgfqpoint{4.393348in}{3.339783in}}{\pgfqpoint{5.000000in}{3.339783in}}%
\pgfpathclose%
\pgfusepath{stroke,fill}%
\end{pgfscope}%
\begin{pgfscope}%
\pgftext[x=2.712500in,y=2.502500in,,base]{\sffamily\fontsize{30.000000}{36.000000}\selectfont 1643}%
\end{pgfscope}%
\begin{pgfscope}%
\pgftext[x=7.287500in,y=2.502500in,,base]{\sffamily\fontsize{30.000000}{36.000000}\selectfont 19}%
\end{pgfscope}%
\begin{pgfscope}%
\pgftext[x=5.000000in,y=6.771033in,,base]{\sffamily\fontsize{30.000000}{36.000000}\selectfont 6640}%
\end{pgfscope}%
\begin{pgfscope}%
\pgftext[x=5.000000in,y=2.502500in,,base]{\sffamily\fontsize{30.000000}{36.000000}\selectfont 23}%
\end{pgfscope}%
\begin{pgfscope}%
\pgftext[x=3.856250in,y=5.171250in,,base]{\sffamily\fontsize{30.000000}{36.000000}\selectfont 6388}%
\end{pgfscope}%
\begin{pgfscope}%
\pgftext[x=6.143750in,y=5.171250in,,base]{\sffamily\fontsize{30.000000}{36.000000}\selectfont 469}%
\end{pgfscope}%
\begin{pgfscope}%
\pgftext[x=5.000000in,y=4.408750in,,base]{\sffamily\fontsize{30.000000}{36.000000}\selectfont 2511}%
\end{pgfscope}%
\begin{pgfscope}%
\pgftext[x=1.568750in,y=1.358750in,,base]{\sffamily\fontsize{30.000000}{36.000000}\selectfont Pathways}%
\end{pgfscope}%
\begin{pgfscope}%
\pgftext[x=8.431250in,y=1.358750in,,base]{\sffamily\fontsize{30.000000}{36.000000}\selectfont Complexes}%
\end{pgfscope}%
\begin{pgfscope}%
\pgftext[x=5.000000in,y=8.372283in,,base]{\sffamily\fontsize{30.000000}{36.000000}\selectfont PPI}%
\end{pgfscope}%
\end{pgfpicture}}
\end{tabular}
\caption{\label{overlap} The overlaps of the protein sets of baker's yeast (left) and human (right). Left: $3,481$ proteins participate in PPIs only, $843$ in PPIs and pathways, $618$ in PPIs, pathways and complexes, $989$ in PPIs and complexes, while $4$ are in pathways only. Right: $6,640$ proteins participate in PPIs only, $6,388$ in PPIs and pathways, $2,511$ in PPIs, pathways and complexes, $469$ in PPIs and complexes, $23$ in complexes and pathways, while $1,643$ are in pathways only and $19$ in complexes only.}
\end{figure}

We compute vectors containing topological information around proteins in the molecular networks: we use graphlets on PPI networks and hypergraphlets on hypergraphs, as described above. To validate our approach, we focus on parts of PPI networks that we know are rich in biological information: protein complexes and pathways. Clearly, not all proteins in a PPI network belong to complexes, or pathways (illustrated in Figure \ref{overlap}). Hence to validate our method, we consider four sets of proteins: those belonging to pathways in human (human-pathways), those belonging to pathways in yeast (yeast-pathways), those belonging to complexes in human (human-complexes), and those belonging to complexes in yeast (yeast-complexes). For each protein in each of these sets, we have two topological signatures: one from the standard graphlets counted on the entire PPI network and one from the hypergraphlet counts in the hypergraph (HG) that  we constructed by using only protein complexes (and equivalently pathways). That is, in an HG, nodes are proteins and each hyperedge represent a protein complex (or pathway) and contains the proteins that belongs to the complex (pathway). For each protein, we also have three biological signatures corresponding to the three levels of GO annotations: BP, MF, and CC. We use these as input into the methods described in Sections \ref{EnrSec} and \ref{CCASec}. The results of these validations are presented in Sections \ref{EnrAn} and \ref{CCAAn}.

The reason for doing these validations on the sets of data for which we know that they are very enriched in biological information (i.e., pathways and complexes) is to demonstrate that our new model and method can correctly identify the biological information. After these validations of the methodology, we use it to perform the analysis of multi-scale protein interaction network data of yeast and human and uncover new biological information. In particular, for each species, we construct a hypergraph that contains all of its PPIs, all of its protein complexes, and all of its pathways; i.e., nodes are proteins and hyperedges correspond to PPIs, protein complexes, and pathways. The results of analysing these hypergraphs with our methods are presented in Section \ref{overlay}.

\begin{figure*}[ht]
\centering
\includegraphics[width=\linewidth]{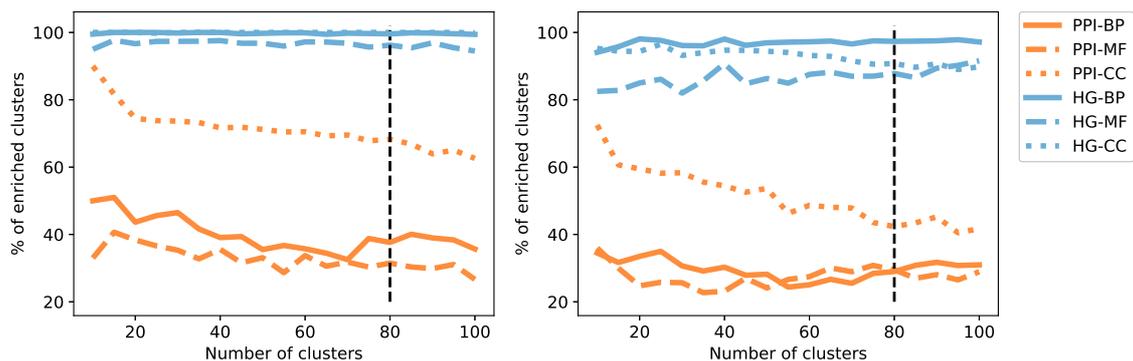}
\vspace{-2em}
\caption{\label{enrichments}The panels give the average percentage of clusters enriched with respect to the total number of clusters for yeast-complexes (left) and yeast-pathways (right), the standard deviation is not represented to avoid overcrowding the panels. The colors represent the models from which the clustering is obtained: HG in blue and PPI in orange. The type of line represents the category of GO annotations: BP are full lines, MF are dashed lines, and CC are dotted line. The black vertical lines signal the number of clusters selected from the set of NMI and SSE curves according to the procedure described in Section \ref{EnrSec}.}
\end{figure*}

\begin{figure*}[ht]
\centering
\scalebox{0.9}{\begin{tabular}{l  c c | c c | c c}
    & \multicolumn{2}{c |}{Biological Process}& \multicolumn{2}{c |}{Molecular Function} & \multicolumn{2}{c}{Cellular Component} \\
        \cline{2-7}
        & HG  & PPI & HG  & PPI & HG  & PPI \\
        \hline
  Yeast-complexes & {\small$100\%$ (51)}  & {\small$53.75\%$ (80)}
  & {\small$100\%$ (49)}  & {\small$51.25\%$ (80)}
  & {\small$100\%$ (51)}  & {\small$74.7\%$ (79)}\\  
  
  Yeast-pathways & {\small$100\%$ (71)}  & {\small $45\%$ (80)}
  & {\small$95.2\%$ (63)}  & {\small$37.5\%$ (80)}
  & {\small $95.4\%$ (65)}  & {\small $56.25\%$ (80)}\\
  
 \hline
 Human-complexes & {\small$94.3\%$ (105)}  & {\small$\textcolor{darkgray}{40.3}\%$ (119)}
 & {\small$82.7\%$ (98)}  & {\small$47.5\%$ (120)}
 & {\small$95.2\%$ (105)}  & {\small$60.8\%$ (120)}\\ 
 
 Human-pathways & {\small$98.2\%$ (111) }  & {\small $59.2\%$ (120)}
 & {\small$98.3\%$ (115)}  & {\small$70.8\%$ (120)}
 & {\small$96.6\%$ (118)}  & {\small $63.3\%$ (120)}\\

 \end{tabular}}
\includegraphics[width=\linewidth]{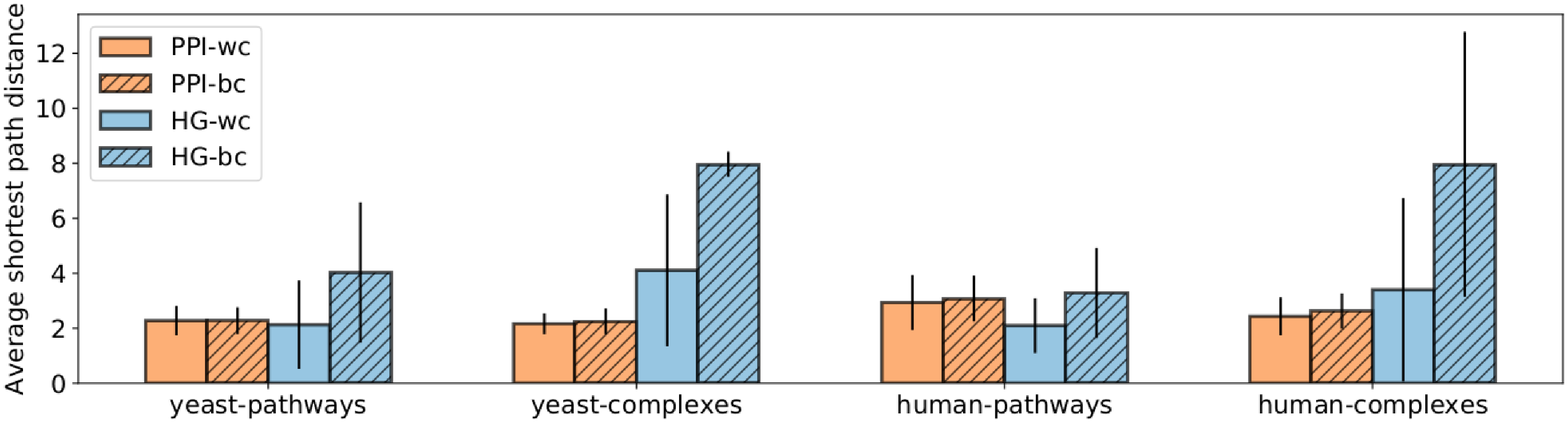}

\scalebox{0.9}{\begin{tabular}{l  c   c  c }
&&&\\
    & Biological Process &Molecular Function & Cellular Component \\
        \hline
  Yeast-complexes & {\small$0.11~(0.0)$ } & 
  {\small$0.1~(0.0)$}& {\small$0.1~(0.0)$}\\  
  
  Yeast-pathways & {\small$0.07~(0.0)$ }
  &{\small $0.07~(0.0)$ }&{\small $0.07~(0.0)$ }\\
  
 \hline
 Human-complexes & {\small$0.07~(0.01)$ }
 & {\small$0.07~(0.02)$}
 & {\small$0.08~(0.06)$ } \\ 
 
 Human-pathways & {\small$0.05~(0.07)$}  
 &{\small$0.06~(0.1)$ } & {\small $0.06~(0.12)$}
 \end{tabular}}

\caption{\label{enrs}  The top table presents the maximum enrichment measured across clusterings obtained with the ``optimal'' number of clusters ($80$ for yeast and $120$ for human). The number in parenthesis is the number of non-empty clusters. The color indicates the statistical significance of the maximum enrichment with respect to random permutation tests: black indicates a significant value, grey a non-significant one. The middle panel gives, for each type of model (HG in blue and PPI in orange), the average of the shortest path lengths within the clusters (wc) and between clusters (bc) of the best clustering obtained for GO--BP annotations. The results are similar for other GO categories and are not presented here due to space limitations. The bottom table presents the results of comparing the  obtained clusterings. We use the HG clustering as baseline and compute the Adjusted Mutual Information (AMI) between the clusterings and the Jaccard Index (in parenthesis) between the sets of enriched GO terms.}
\end{figure*}

\section{Results \& Discussion}

\subsection{Validation of our methodology}
\subsubsection{Enrichment Analysis \label{EnrAn}}

Having computed the topological vectors from both network models (PPI and HG) for each protein of each of the four sets of proteins described in Section \ref{stuff} (human-pathways, human-complexes, yeast--pathways and yeast--complexes), we apply the methodology detailed in Section \ref{EnrSec} to investigate if similarly wired proteins have similar functions. Interestingly, the percentage of enriched clusters is relatively stable as we increase the number of clusters. Hence, any partitioning of the proteins based on the local wiring patterns in a network, quantified by using graphlets or hypergraphlets, captures the underlying biological information (see Figure \ref{enrichments}). This underlines the crucial role played by the way proteins interact in determining protein function without any information about their sequence, or interacting partners. Furthermore, when examining the clusterings obtained at a specific number of clusters, $k$ (see Section \ref{EnrSec} for details on how $k$ is chosen), we observe that the enrichments (top table in Figure \ref{enrs}) are all statistically significant, except for the one in gray. Importantly, clusters obtained from HG models are more enriched than those obtained from PPI networks. This result validates the relevance of our HG modeling in capturing the underlying biological information and underlines the potential of hypergraphlets for mining molecular hypernetworks.

To further investigate the clusterings, we compute for each the average shortest path distances between pairs of proteins belonging to the same clusters (``within-clusters'') and between pairs of proteins which are in different clusters (``between-clusters''; see  middle panel in Figure \ref{enrs}). We observe a larger gap between within-cluster and between-clusters average shortest path lengths for clustering obtained from higher order molecular organisation than from clusterings obtained from PPI networks. Hence, proteins that are topologically similar in the HG model in addition to sharing biological functions tend to be at shorter distance from each other. This result is consistent with the literature on ``guilt by associations'', which predicts protein functions from their neighbourhoods in molecular networks \cite{vazquez2003global}.

Finally, we observe that the clusterings obtained from the PPI model are different from those obtained from the HG model both in terms of GO annotations that are enriched and in terms of clustered proteins (see bottom table in Figure \ref{enrs}). This is because a Jaccard Index close to $0$ means that the sets of the enriched GO terms in the PPI and HG clusterings tend not to overlap. Also, AMI scores below $0.1$  mean that pairs of proteins belonging to the same clusters in one clustering are typically in different clusters in the other clustering. This demonstrates that modeling the interactomes by hypergraphs will uncover new biological information that cannot be uncovered from the analysis of PPI networks. Also, it demonstrates the complementarity of the two representations and that the two are capturing different underlying biological information. 

\begin{figure}[ht]
\centering
\includegraphics[width=0.5\linewidth]{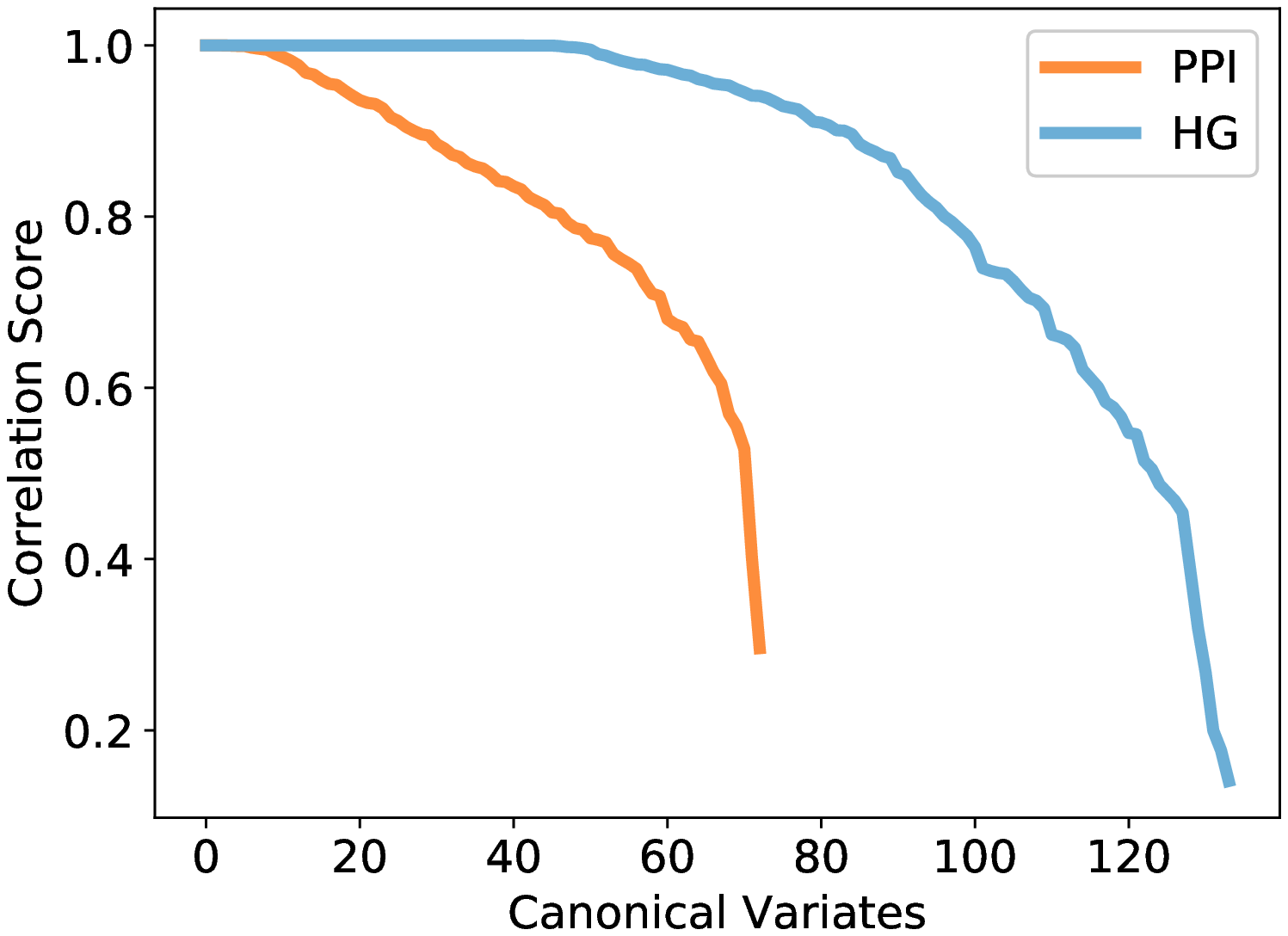}\includegraphics[width=0.5\linewidth]{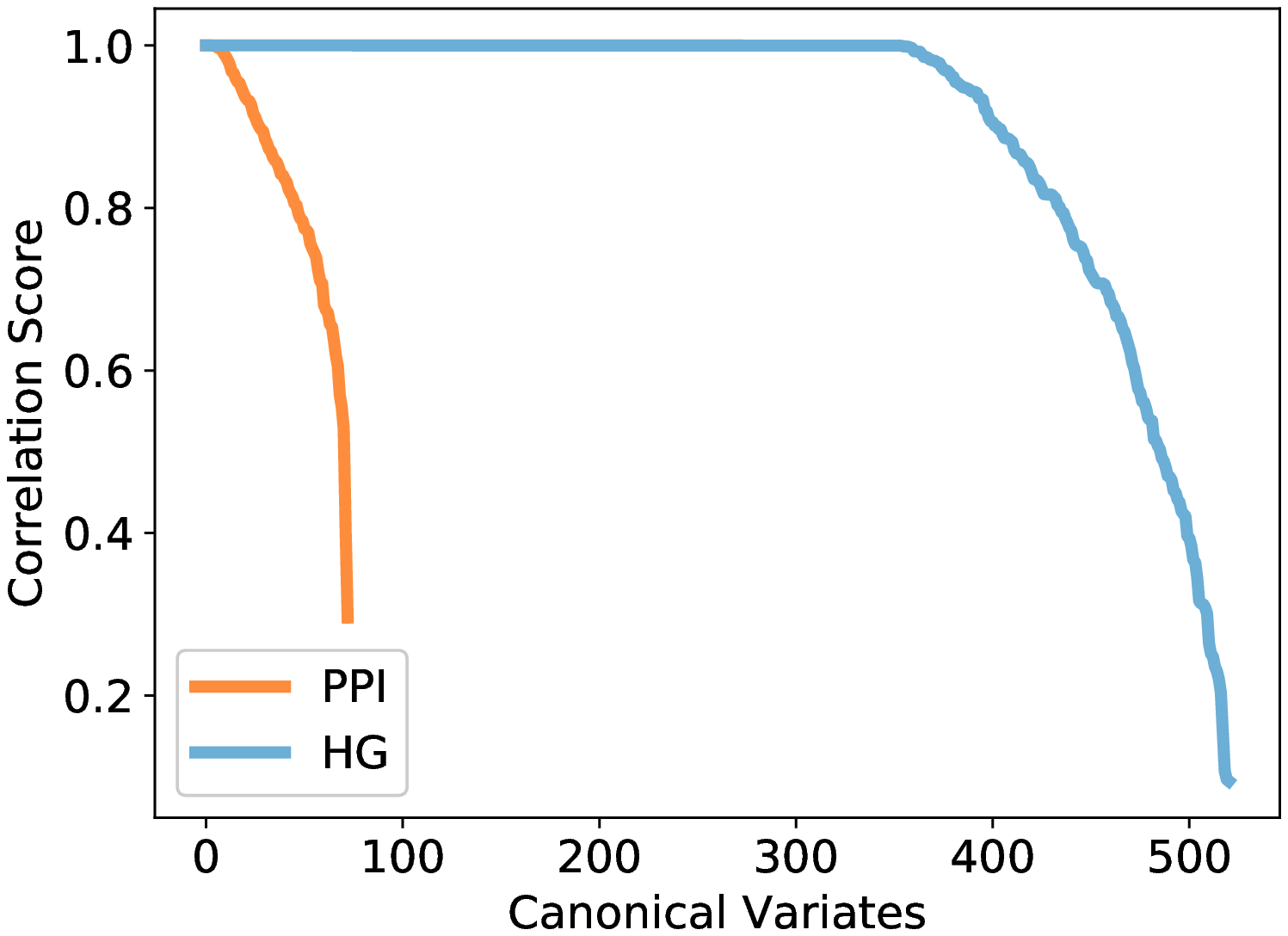}
\caption{\label{cca} Canonical correlation score distribution for yeast-complexes (left) and yeast-pathways (right). The canonical variates represented are all statistically significant (p-value $\leq 5\%$) and are sorted by correlation score. The colors represent the model and the topological signatures from which the canonical variates are obtained: HG in blue and PPI in orange.}
\end{figure}

\subsubsection{Canonical Correlation Analysis \label{CCAAn}}

We investigate the existence of specific topology-function links, i.e. the connection between specific hypergraphlets (or graphlets) and GO annotations by using CCA described in Section \ref{CCASec}. We apply it on the same PPI and HG of yeast and human used in the clustering and enrichment analysis (Section \ref{EnrAn}): for each set of proteins, we compute the CCA between the topology-containing vectors of each of the associated models (PPI and HG) and the vector of GO annotations for each category (BP, MF, and CC). Due to space limitations, we present only the results obtained for yeast and GO--BP annotations. We obtain similar results in all other cases and the discussion below holds for them as well.

We observe that each model has a number of canonical variates with correlation close to $1$ (Figure \ref{cca}), which indicates a strong topology-function relationship in these data that was previously highlighted in the context of economic network data \cite{Yaveroglu2014}. In particular, this means that some functions are strongly linked to specific wiring patterns and thus, local topology can potentially be used for predicting protein functions. For that purpose, hypergraphlets of HGs have a strong advantage over graphlets of PPI networks in the number of canonical variates with a score close to $1$, which is $3$ to $13$ times more variates with HGs. This is also expected, since we chose our hypernetworks to model already function rich parts of molecular networks, protein complexes and pathways, and it validates our methodology.

\begin{figure}[ht]
\centering
\includegraphics[width=0.7\linewidth]{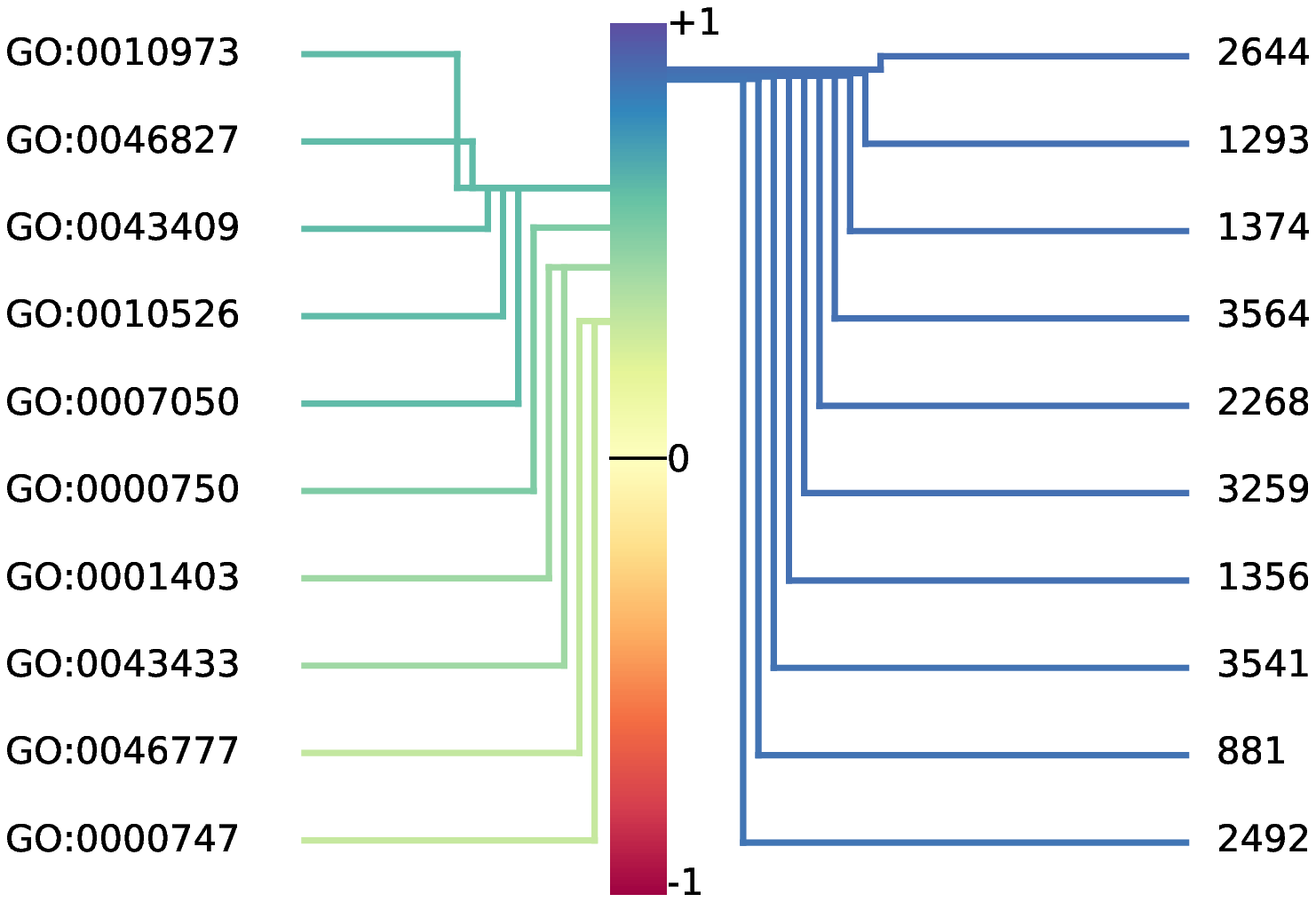}

\begin{tabular}{r p{5cm}}
&\\
{\color{darkblue} GO:0010973:}& {\textcolor{darkblue}{positive regulation of barrier septum assembly}}\\
{\color{darkblue} GO:0046827:}& { 
\textcolor{darkblue}{positive regulation of protein export from nucleus}}\\
{\color{darkblue} GO:0043409:}& { \textcolor{darkblue}{negative regulation of MAPK cascade}}\\
{\color{darkblue} GO:0010526:}& { \textcolor{darkblue}{negative regulation of transposition, RNA-mediated}}\\
{\color{darkblue}GO:0046777:}& {\textcolor{darkblue}{protein autophosphorylation}}\\
{ GO:0007050:}& {cell cycle arrest}\\
{\color{darkblue}GO:0000750:} & {\textcolor{darkblue}{pheromone-dependent signal transduction involved in conjugation with cellular fusion}}\\
{\color{darkblue}GO:0001403:}& {\textcolor{darkblue}{invasive growth in response to glucose limitation}}\\
{\color{darkblue}GO:0043433:}& {\textcolor{darkblue}{negative regulation of sequence-specific DNA binding transcription factor activity}}\\
{\color{darkblue}GO:0000747:}& {\textcolor{darkblue}{conjugation with cellular fusion}}\\

\end{tabular}\scalebox{0.65}{\begin{tabular}{c c c}
\includegraphics[width=0.2\linewidth]{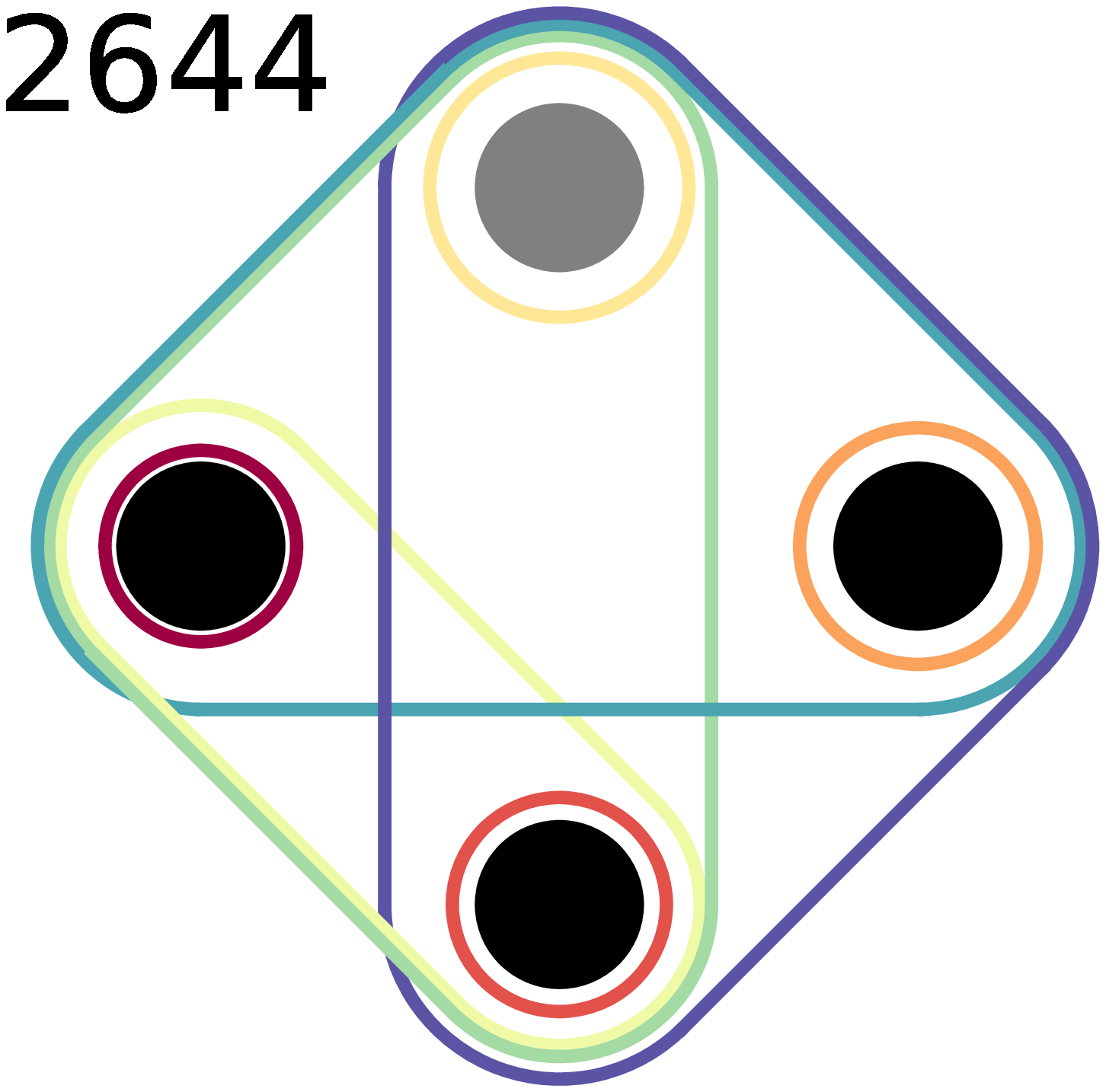} &
\includegraphics[width=0.2\linewidth]{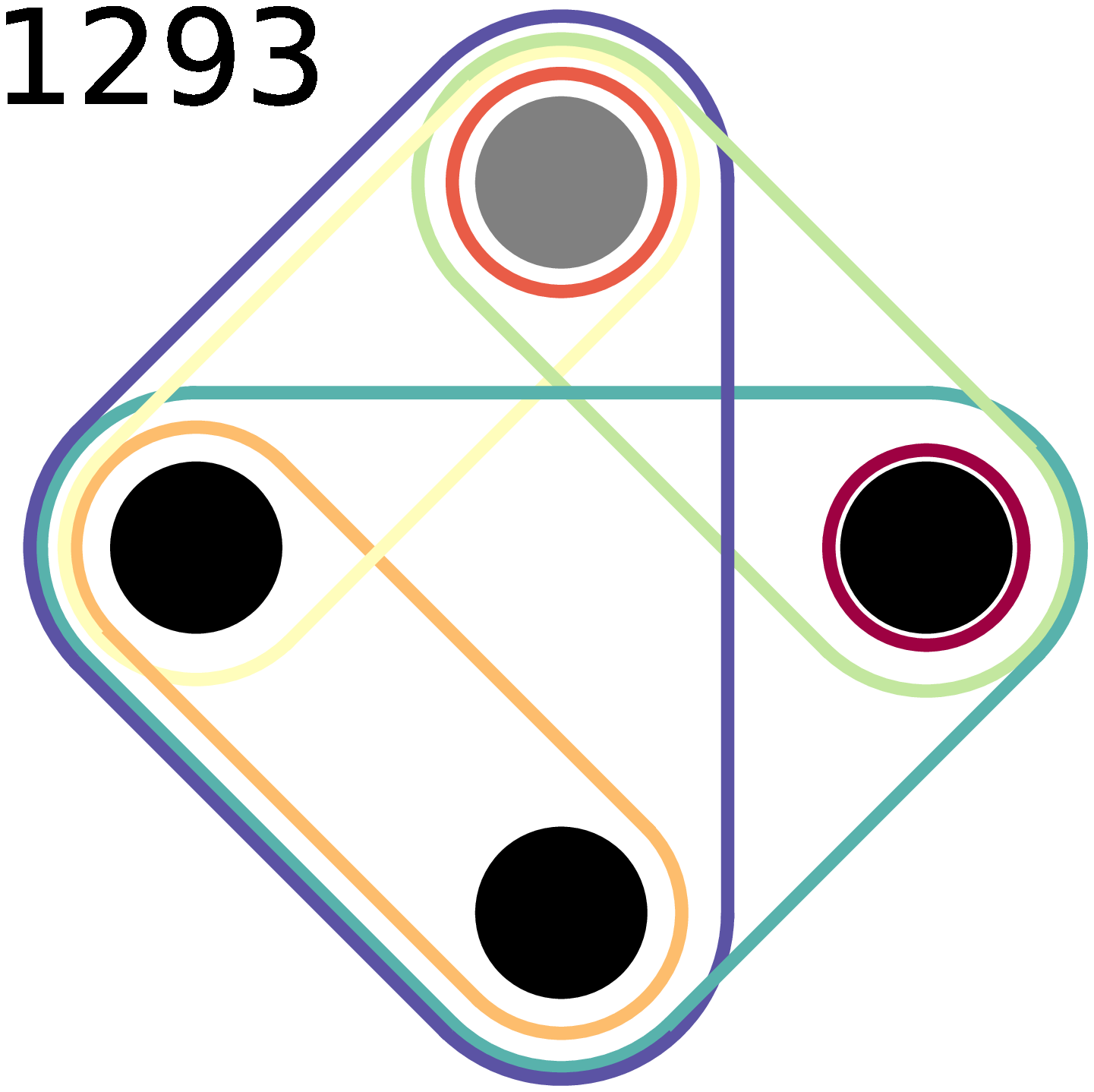} &
\includegraphics[width=0.2\linewidth]{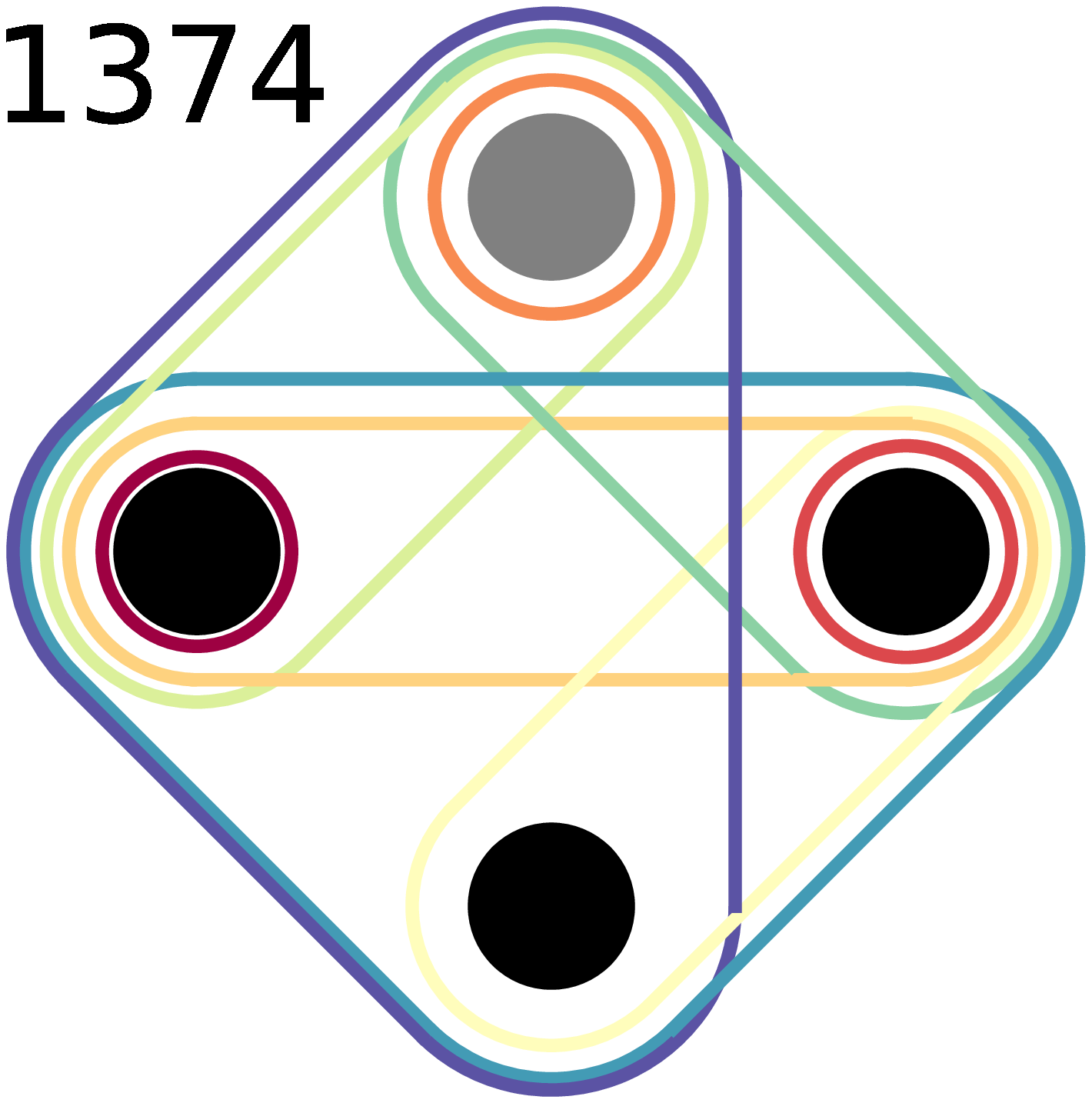} \\
\includegraphics[width=0.2\linewidth]{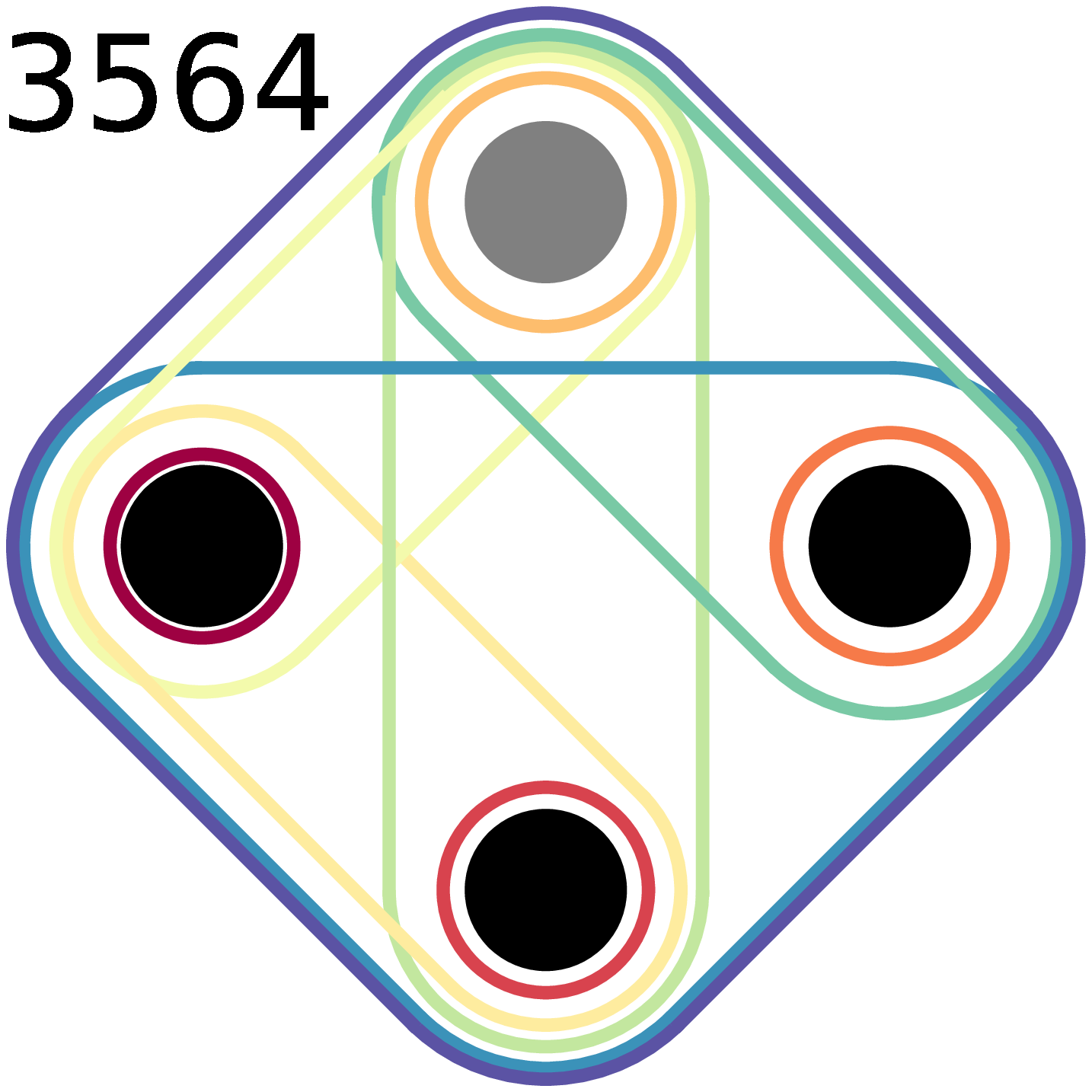} &
\includegraphics[width=0.2\linewidth]{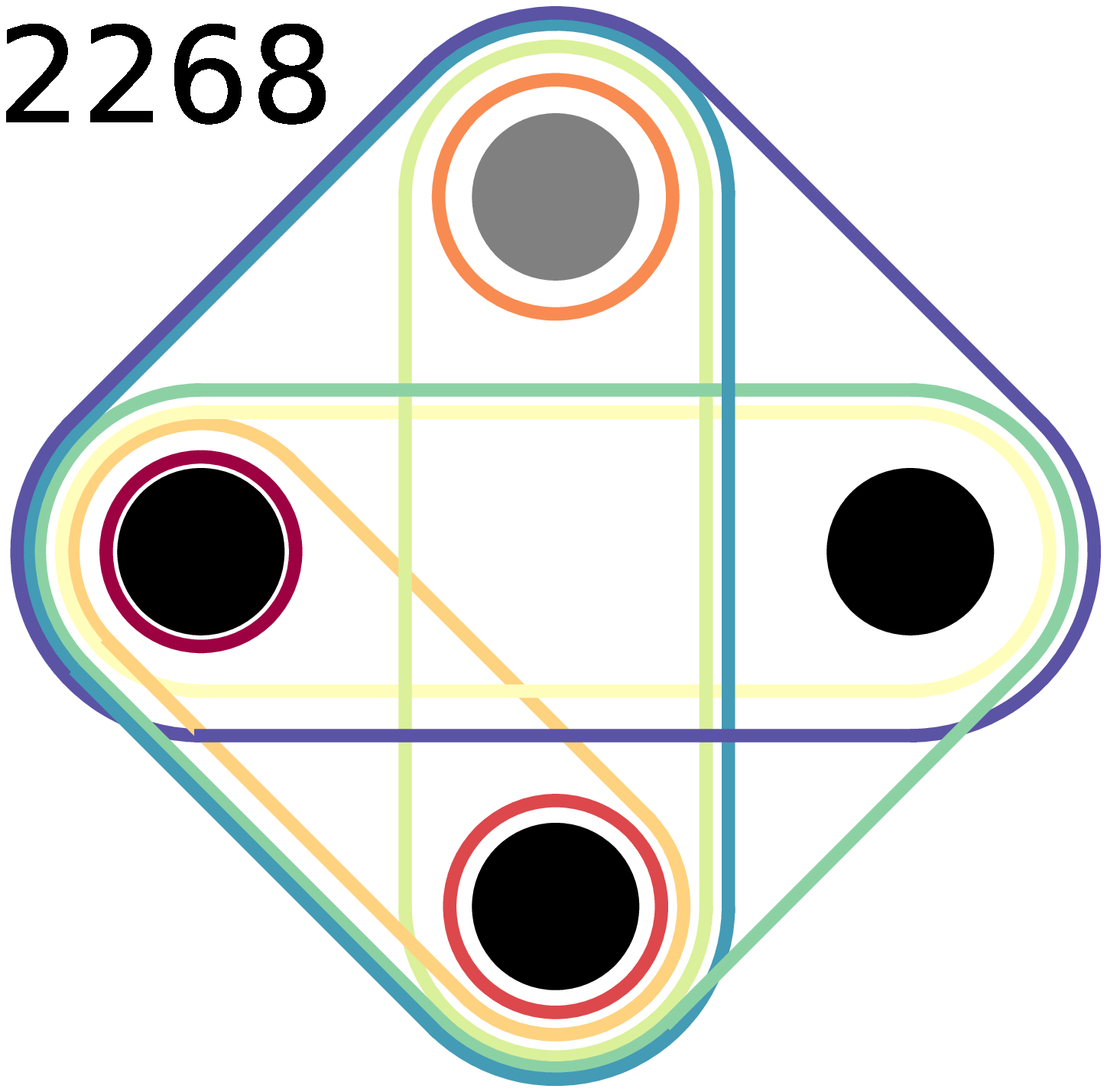} &
\includegraphics[width=0.2\linewidth]{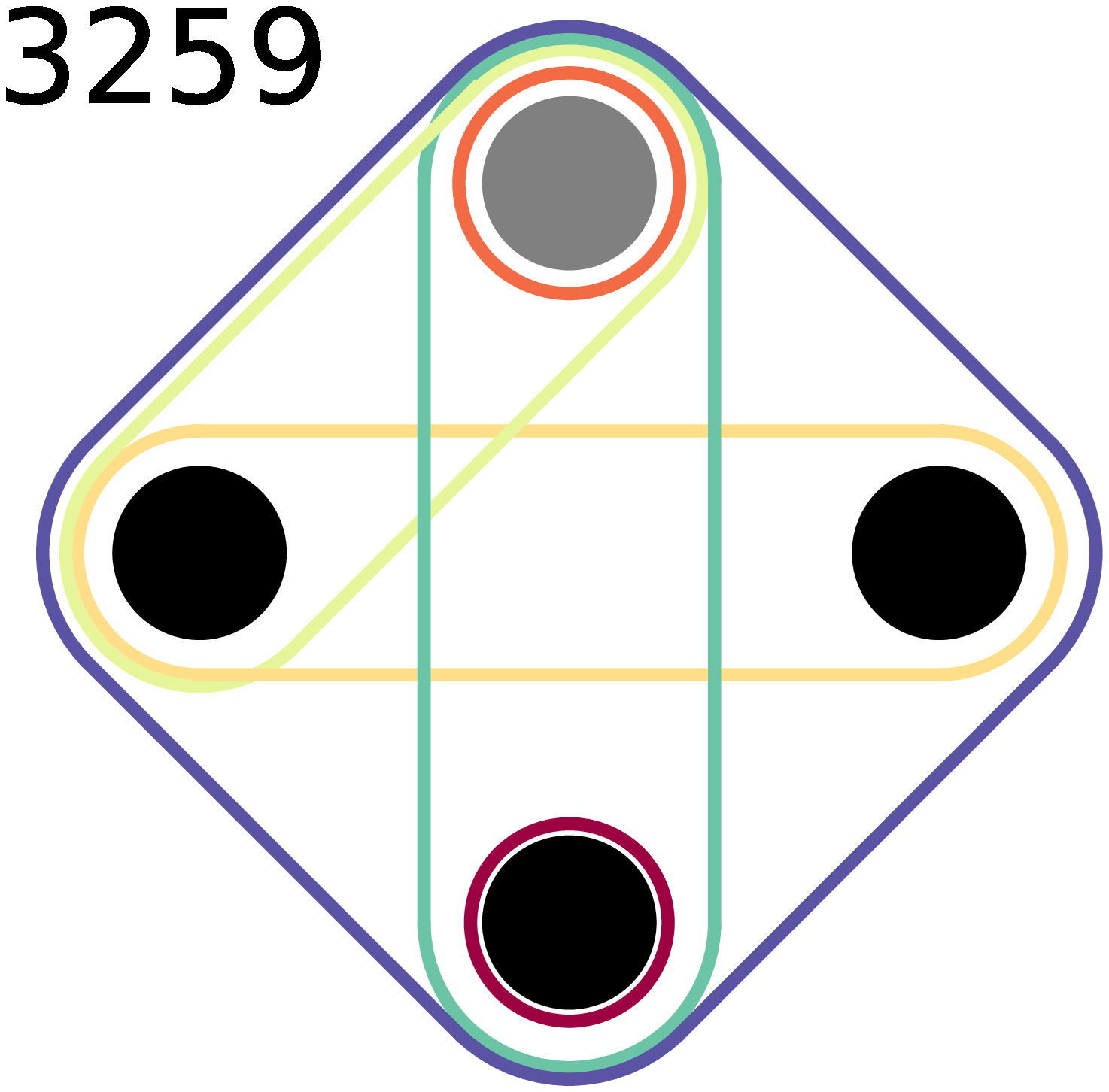} \\
\includegraphics[width=0.2\linewidth]{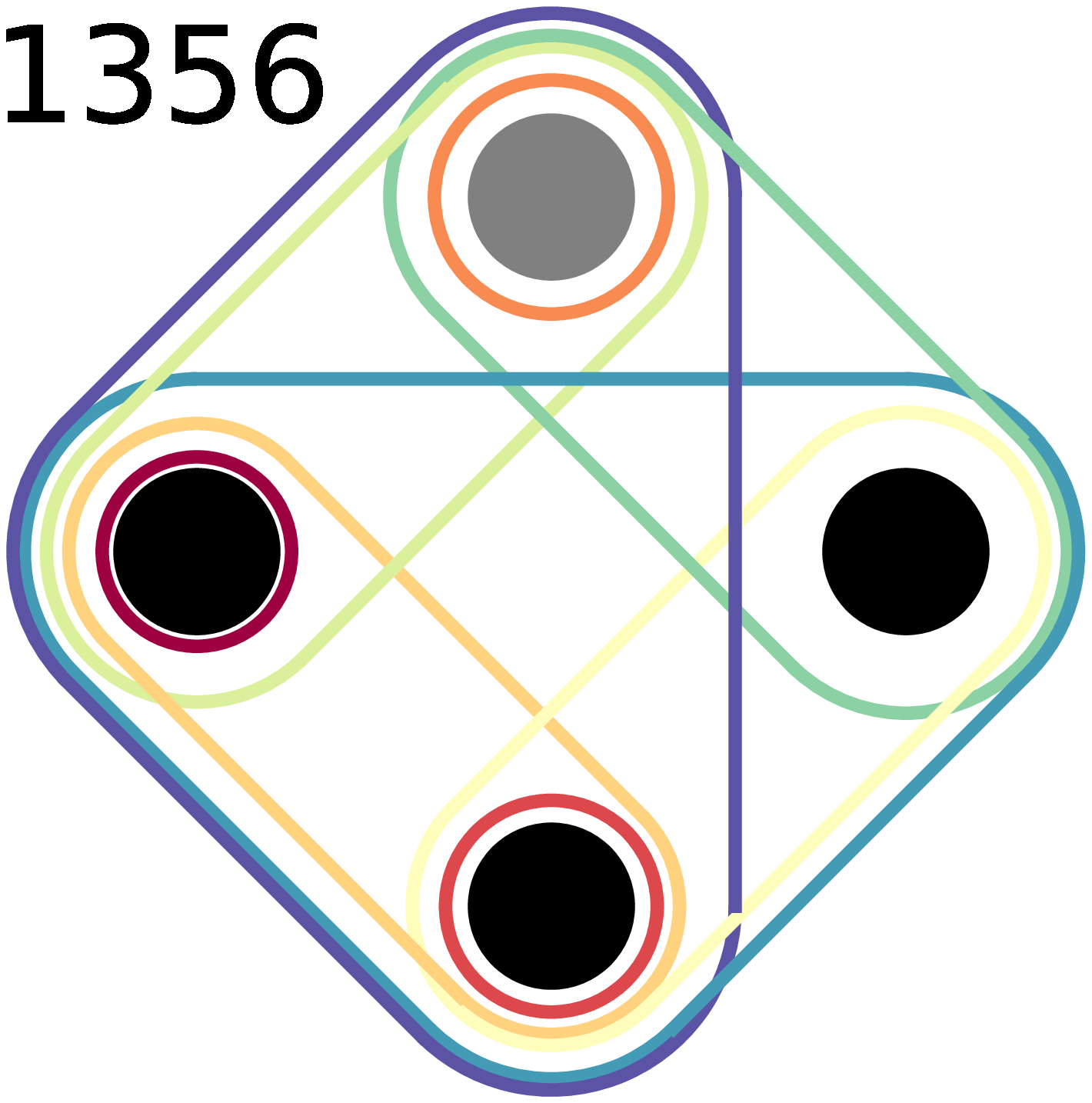} &
\includegraphics[width=0.2\linewidth]{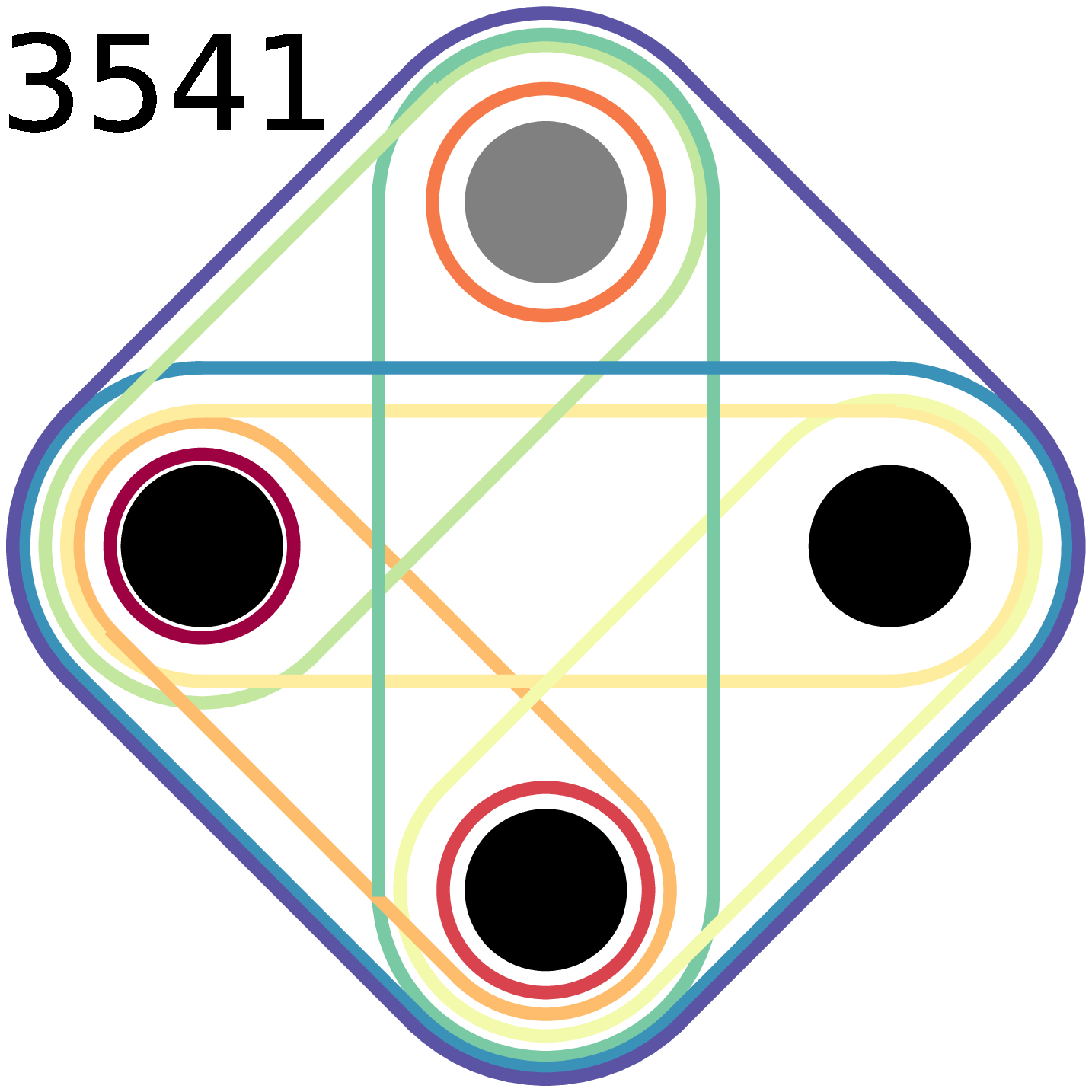} &
\includegraphics[width=0.2\linewidth]{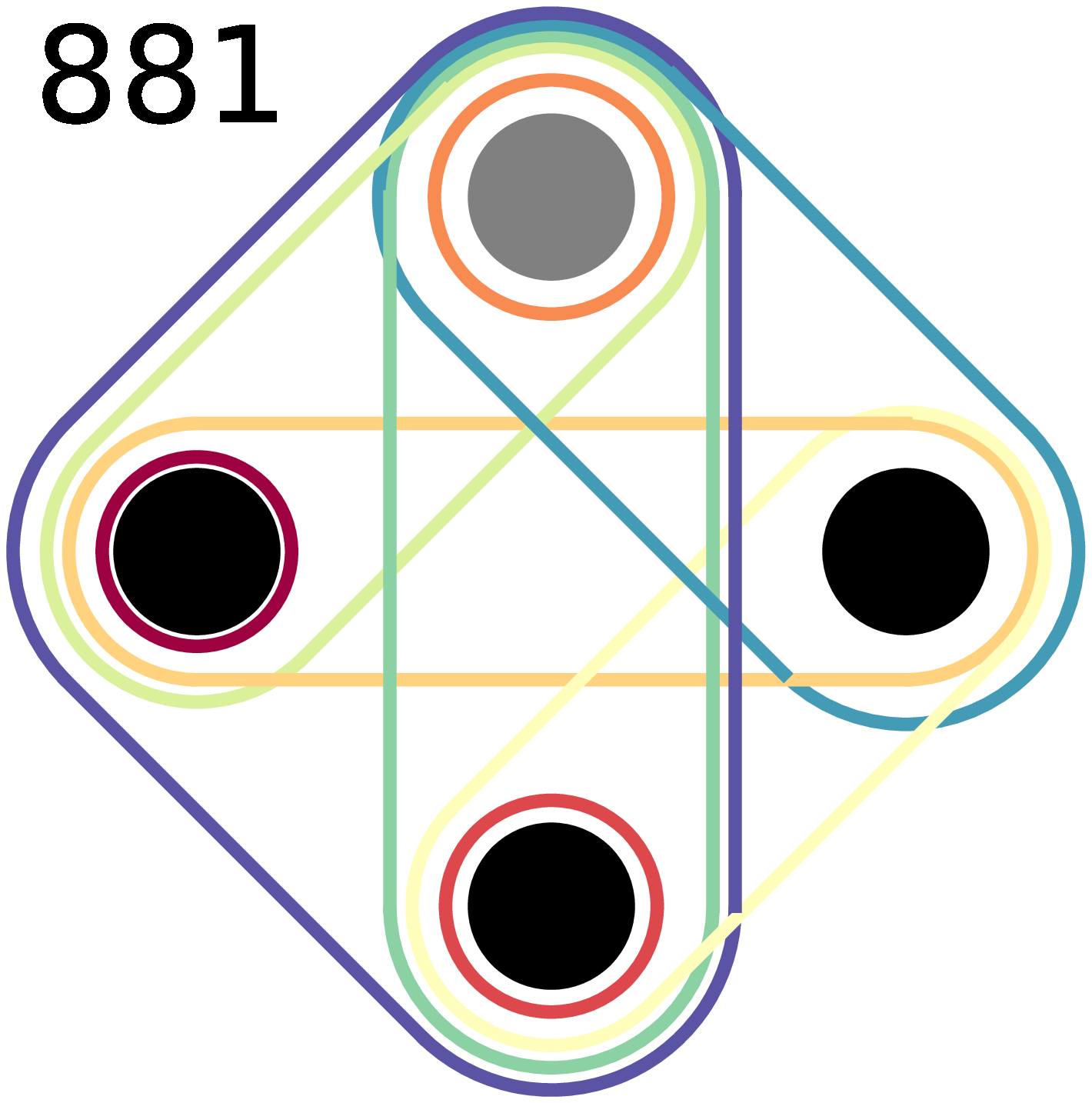} \\
\includegraphics[width=0.2\linewidth]{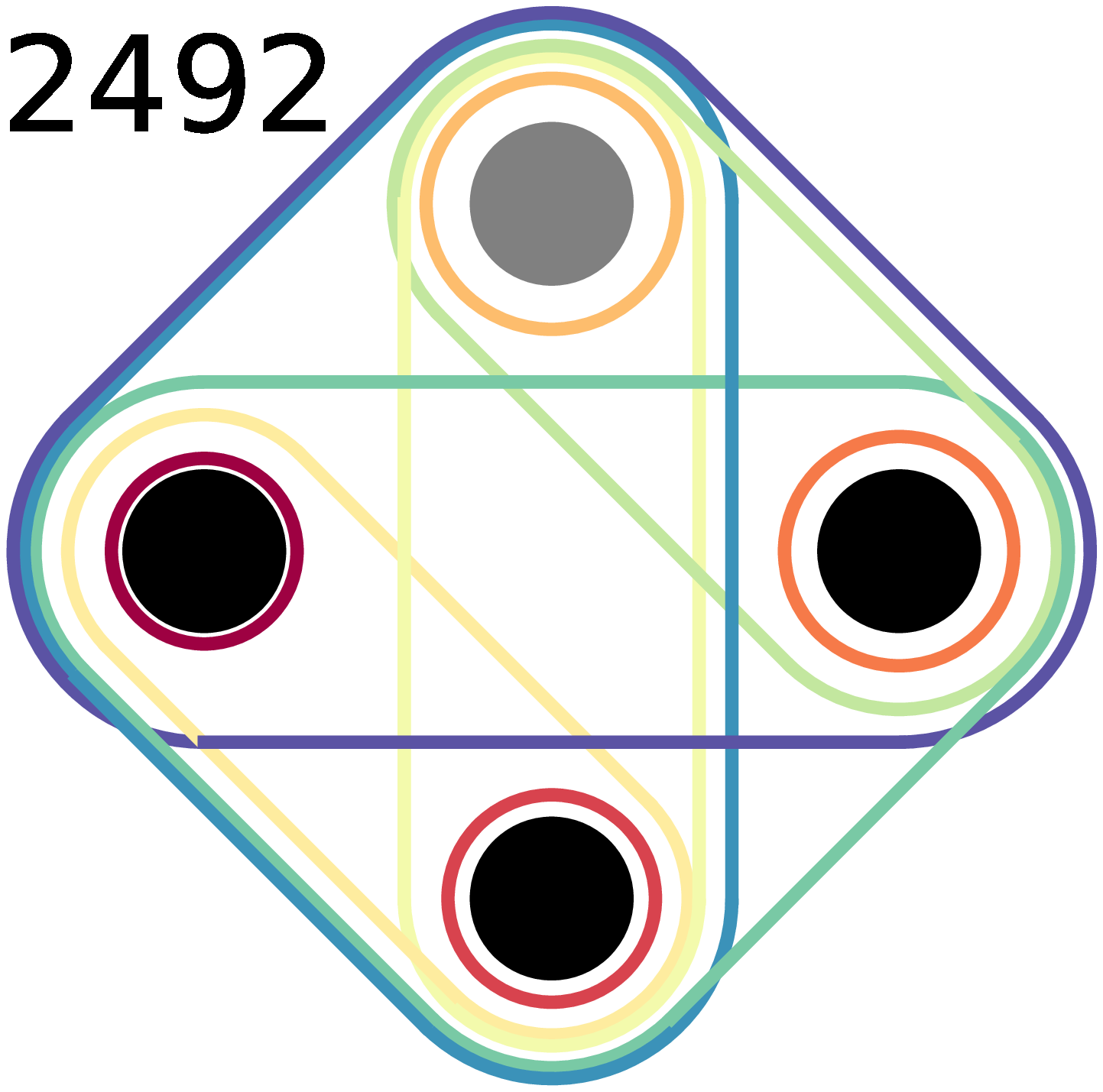} 
\end{tabular}}

\caption{\label{variates} The most significant CCA variate between HDVs of the proteins of yeast-pathways and their GO--BP annotations. The correlation score between the linear combination of annotations and the linear combination of hypergraphlet orbits is $1$. The annotations (orbits) illustrated above correspond to the $10$ that have the highest Pearson's correlation scores with respect to the linear combinations of annotations (orbits). Each GO term in blue font is annotating at least one protein conjointly with at least one other annotation that is also denoted in blue font, according to QuickGO ontology search engine \cite{binns2009quickgo}.}
\end{figure}

In Figure \ref{variates}, we take a closer look at the most significant CCA variate. The variate score of $1.0$ links a linear combination of GO annotations to a linear combination of hypergraphlets orbits.  For instance, this means that a gene annotated with positive regulation of barrier spectrum assembly (GO:0010973) will likely have a relatively large $2644^{th}$ orbit degree  in the hypernetwork. Why these specific orbits are linked to these functions is a question that is outside of the scope of this study and that needs to be further investigated. We find that the GO terms identified here are also biologically coherent: each of the GO--BP terms denoted in blue text in Figure \ref{variates} is annotating at least one protein conjointly with at least one other annotation, that is also denoted in blue text in Figure \ref{variates}, according to QuickGO search engine \cite{binns2009quickgo}. Furthermore, the only remaining annotation, cell cycle arrest (GO:0007050), has been linked to the MAPK pathway in the literature \cite{pumiglia1997cell}, as have been most of the other terms \cite{madhani1998control,gustin1998map}. Hence, the entire set of GO annotations presented in Figure \ref{variates} is biologically coherent, which validates the relevance of the canonical variate and of our hypergraph-based methodology in capturing functional information.

\subsection{Analysing multi-scale molecular organisation \label{overlay}}
To explicitly capture the multi-scale organisation of protein interactions, we model them by a hypernetwork containing all PPIs, all protein complexes and all biological pathways as hyperedges (detailed in Section \ref{stuff}).
To assess if the wiring patterns in our new HG model capture the biological functions of proteins, we do the clustering and enrichment analysis (Section \ref{EnrSec}), as well as the canonical correlation analysis (Section \ref{CCASec}) on these hypernetworks of baker's yeast and human. We compare the results with those that we obtain by applying the same methodologies to PPI networks. 

\begin{figure}[ht]
\centering
\includegraphics[width=\linewidth]{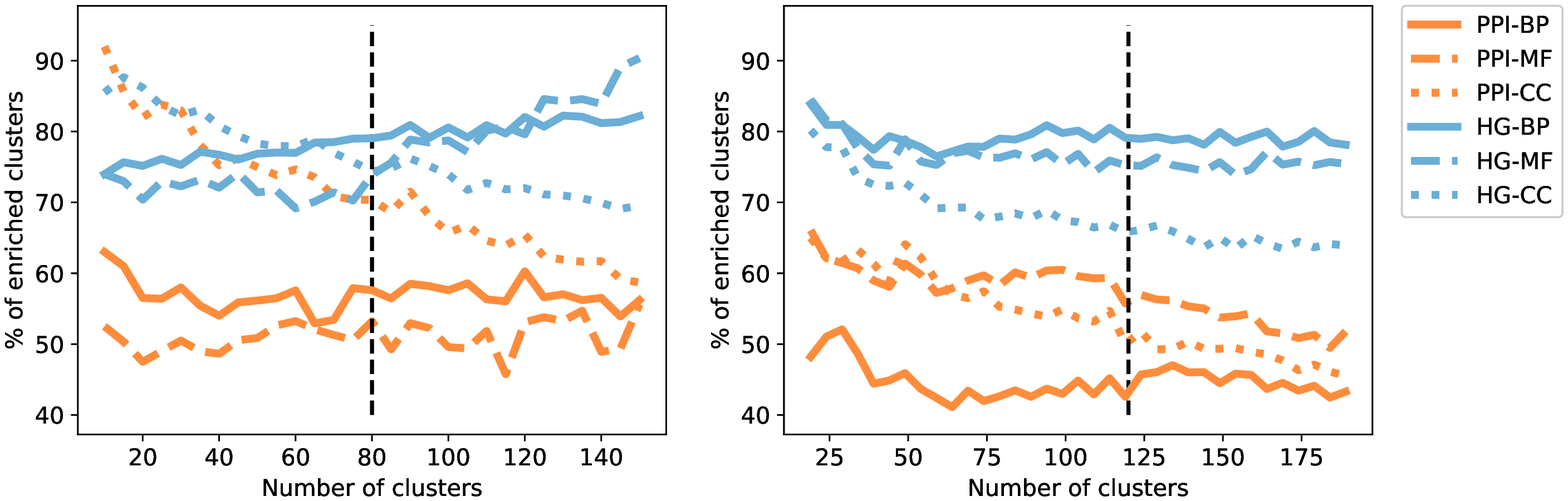}

\begin{tabular}{l  c c | c c | c c}
    & \multicolumn{2}{c |}{Biological Process}  & \multicolumn{2}{c |}{Molecular Function} & \multicolumn{2}{c}{Cellular Component}\\
        \cline{2-7}
        & HG  & PPI & HG   & PPI & HG  & PPI \\
        \hline
  Yeast & {\small$91.1\%$ (79)}  & {\small$68.75\%$ (80)}
  & {\small$71.8\%$ (79)}  & {\small$52.5\%$ (80)}
  & {\small$88.6\%$ (79)}  & {\small$72.2\%$ (79)}\\  
  
 \hline
 Human & {\small$100.0\%$ (105)}  & {\small$41.7\%$ (120)}
 & {\small$95.0\%$ (98)}  & {\small$55.8\%$ (120)}
 & {\small$76.1\%$ (105)}  & {\small$51.7\%$ (120)}\\ 
 \end{tabular} 
 
 \includegraphics[width=\linewidth]{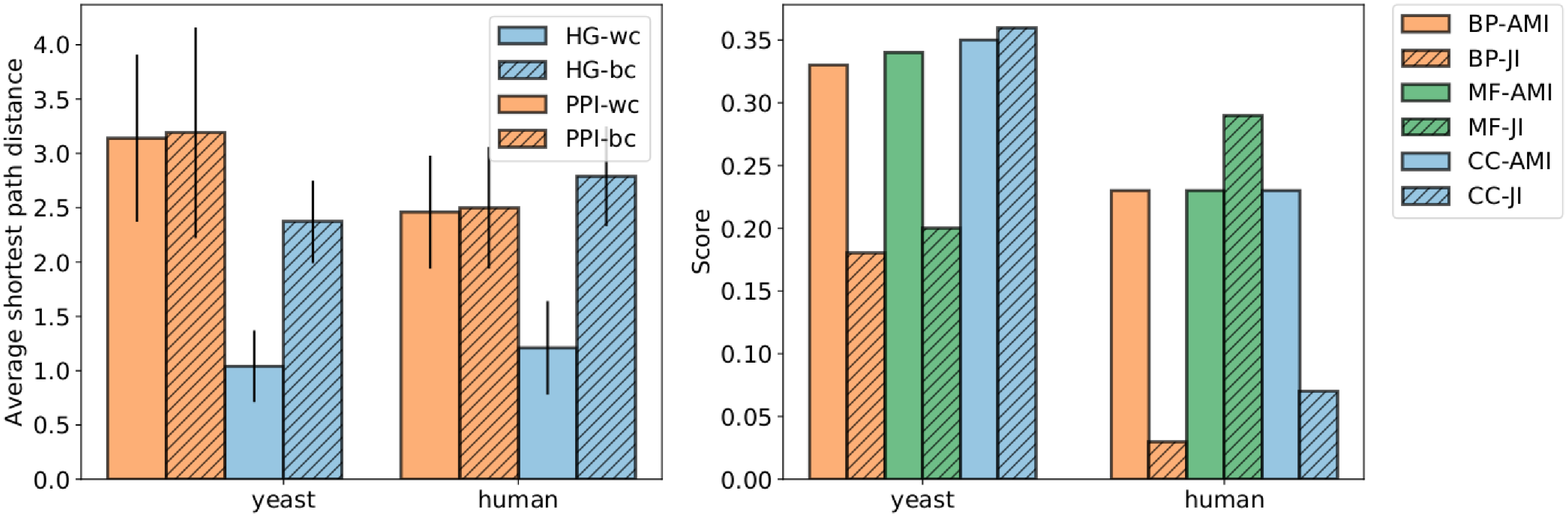}
 
\caption{\label{hyperenr} The top panels give the average percentages of clusters enriched with respect to the total number of clusters for yeast (left) and human (right), the standard deviation is not represented to avoid overcrowding the panels. The colors represent the models from which the clustering is obtained: HG in blue and PPI in orange. The type of line represents the category of GO annotations: BP are full lines, MF are dashed lines, and CC are dotted line. The black vertical lines denote the number of clusters selected from the set of NMI and SSE curves according to the procedure described in Section \ref{EnrSec}. The middle table presents the maximum enrichment measured across clusterings obtained with the ``optimal'' number of clusters (denoted by the black vertical lines in the top panels). The number in parenthesis is the number of non-empty clusters. All enrichments are significant. The bottom left panel gives, for each type of model (HG in blue and PPI in orange), the average of the shortest path lengths within the clusters (wc) and between clusters (bc) of the best clustering obtained for GO--BP annotations. The results are similar for other GO categories and are not presented here due to space limitations. The bottom right panel represents the results of the comparison of the  obtained clusterings. We use the HG clustering as baseline and compute the Adjusted Mutual Information (AMI) between the clusterings and the Jaccard Index (JI) between the sets of enriched GO terms. }
\end{figure}

In these unifying HG models of multi-scale molecular organisation, we observe that clusterings of the proteins based on their topological vectors in a network, obtained by using graphlets or hypergraphlets, capture the underlying biological information (see the top panels of Figure \ref{hyperenr}). Furthermore, the clusters obtained from the hypernetwork topology lead to higher enrichments in GO--BP, GO--MF, and GO--CC annotations. This shows that our newly proposed model, regardless of the choice of the total number of clusters, $k$, captures more protein biological function in its topology than the standard PPI networks.

When choosing the number of clusters, $k$, according to the criteria detailed in Section \ref{EnrSec}, we observe that all enrichments are statistically significant and that the HG models allow for an increase of over $15\%$ in the number of enriched clusters when compared to the PPI networks. This finding underlines the link between multi-scale interaction patterns and biological functions. Interestingly, when investigating the clusters, we observe that a majority of the proteins in the non-enriched clusters only have reported PPIs, but not any pathways or complexes that they belong to. This is true for $59\%$ of the proteins in the HG model of yeast and $38\%$ of the proteins in the HG model of human. This might be due to incompleteness of the pathways and protein complexes data. Our results indicate that when more complete data on complexes and pathways becomes available, our methodology will be able to extract additional biological information.

We observe that proteins clustered using topological features derived from representations of multi-scale molecular organisation tend to also be closer in terms of shortest path distances compared to those obtained by clusterings based on the topology of PPI networks (see bottom left panel in Figure \ref{hyperenr}). Interestingly, most proteins clustered together in the HG models are direct neighbours or second neighbours. Hence, the fact that we obtain enriched biological functions in those clusters is consistent with empirical evidences showing that $70$-$80\%$ of interacting proteins share at least one function. Those evidences were the motivation for the \textit{majority rule} used in the literature for functional prediction \cite{vazquez2003global}.

Finally, we observe  that the clusterings obtained from the PPI models are different from those obtained from the HG models both in terms of GO annotations that are enriched, with a Jaccard Index below $0.25$, and in terms of similarity of clusters, with an AMI below $0.35$ (see bottom right panel in Figure \ref{hyperenr}). This confirms that our multi-scale model is not equivalent to the standard PPI network and uncover additional biological information complementary to that of the PPI network.

\begin{figure}[ht]
\centering
\includegraphics[width=0.7\linewidth]{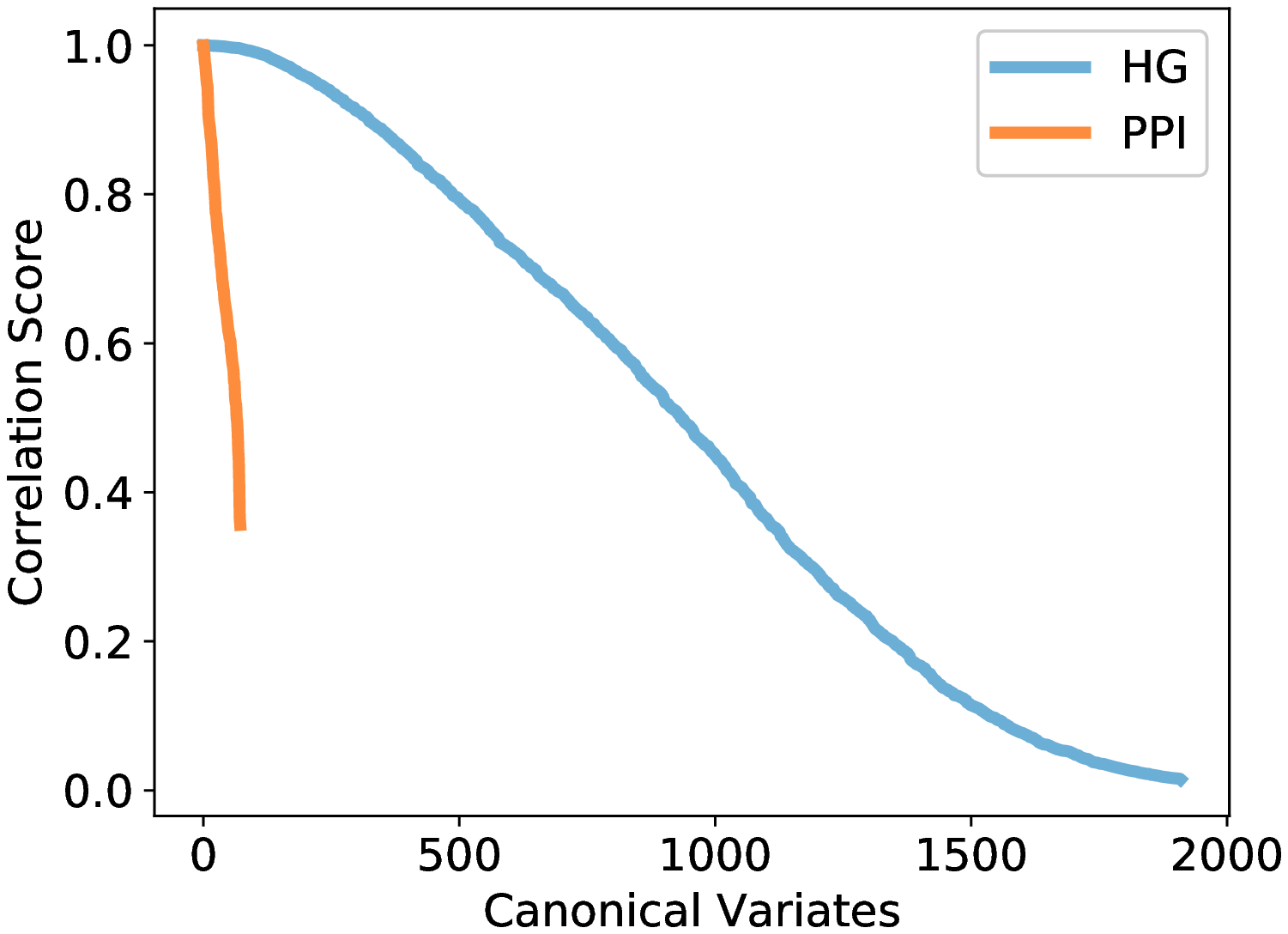}
\caption{\label{hypercca} Canonical correlation score distribution for yeast-complexes (top) and yeast-pathways (bottom). The canonical variates represented are all statistically significant (p-value $\leq 5\%$) and are sorted by correlation score. The colors represent the model and the topological signatures from which the canonical variates are obtained: HG in blue and PPI in orange. }
\end{figure}

Using CCA (Section \ref{CCASec}), we observe that each model has high scoring canonical variates, which indicates that some functions are strongly linked to specific wiring patterns (see Figure \ref{hypercca}). For that purpose, hypergraphlets of our new HG models have an advantage over graphlets of PPI networks in the number of canonical variates with high correlation score: it has over $300$ canonical variates with score greater than $0.9$ compared to only $10$ for PPI networks.
This indicates that the HG model's local wiring patterns are more correlated with the underlying biology that those of the PPI networks.

Finally, we use the clusterings to investigate the potential of our newly proposed models in conjunction with our hypergraphlets to predict protein functions. As demonstrated above, we identified clusters of proteins with significantly enriched GO annotations. We use these clusters to predict the functions of proteins. For each GO category, we identify two disjoint sets of proteins in each of our hypernetworks: the set of proteins that are experimentally annotated with at least one of the enriched GO terms in their cluster (on which the enrichment computations are based) and the set of proteins that have some predicted annotation in the GO database.

\begin{figure}[ht]
\centering
\includegraphics[width=0.7\linewidth]{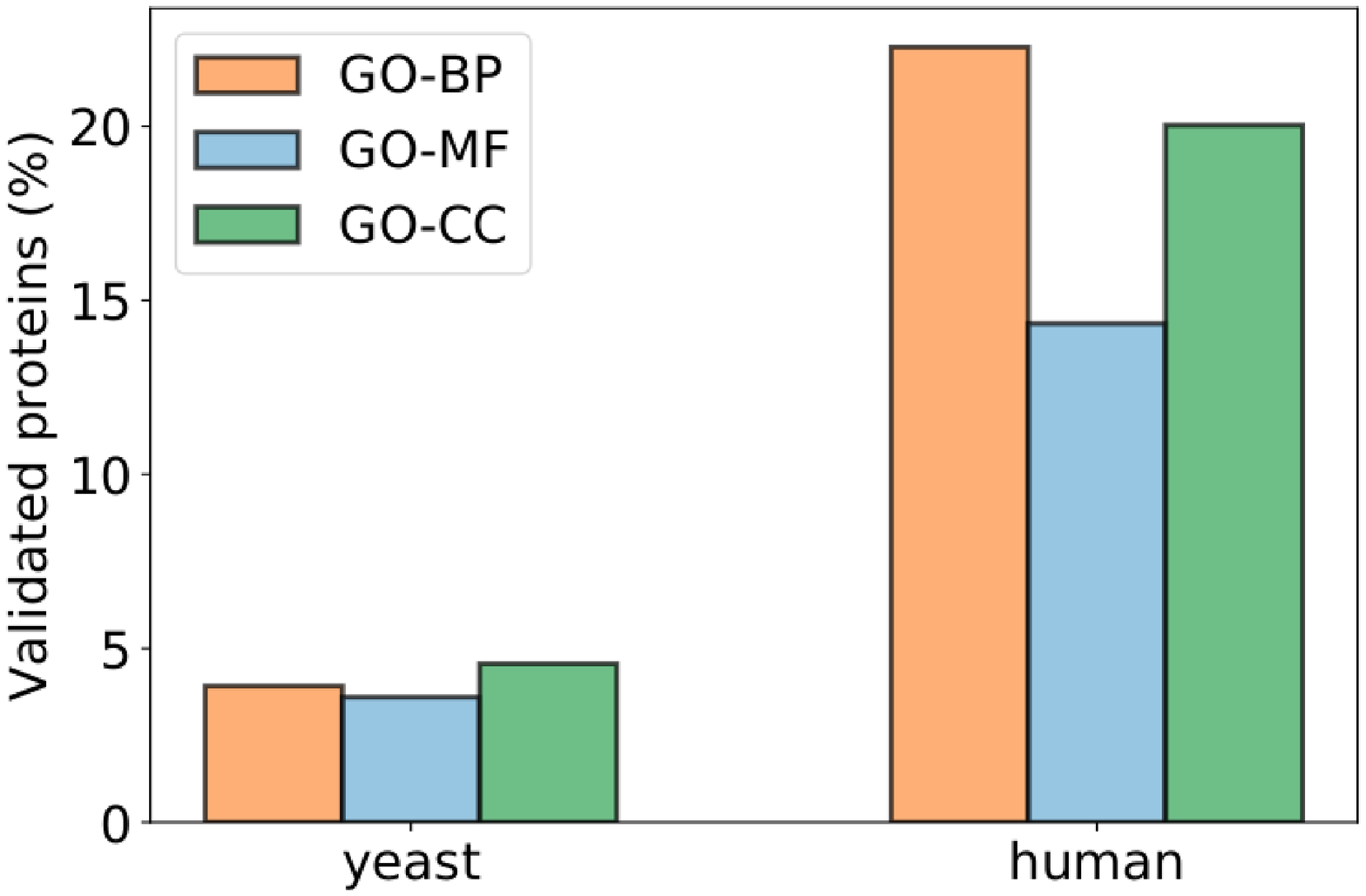}
\caption{\label{preds} Percentages of proteins that have at least one of the enriched terms of their clusters in their set of predicted GO annotations (obtained from the GO database \cite{Consortium2015}). The values correspond to the number of such proteins out of the number of proteins that have at least one putative annotation in the GO database and are not experimentally annotated with any of the enriched terms of their clusters.}
\end{figure}

First, we consider the second set and investigate how many of those proteins have at least one of the enriched terms of their cluster as their predicted GO annotation \cite{Consortium2015}. For GO--BP, this set contains $11,686$ proteins for human ($4,161$ for yeast). For GO--MF, it contains $7,243$ proteins for human ($3,586$ for yeast). For GO-CC, it contains $6,589$ proteins for human ($3,510$ for yeast). We show that out of these proteins, about $5\%$ for yeast and $15$--$23\%$ for human have been putatively annotated in GO with at least one of our enriched functions in their clusters (see Figure \ref{preds}), which validates our approach.

Second, we focus on the proteins of the hypernetworks that are unannotated in GO database (this corresponds to $994$ proteins for human and $97$ proteins for yeast) and investigate the GO--BP annotations we predict for them. We predict function for each of these proteins by associating it with the enriched experimentally obtained GO term that annotates the most proteins in its cluster. We survey the literature to validate our top predictions for human (the top predictions correspond to the most statistically significantly enriched GO terms). We predict that HIST1H2AJ is involved in nucleosome assembly (GO:0006334), which is confirmed in the literature \cite{diaz1996prothymosin}. We further predict that XIST is linked to chromatin organization (GO:0006325), which has also been highlighted in past studies \cite{brockdorff1992product}. We also predict that NME1-NME2 (an unknown protein encoded between NME1 and NME2 in the DNA)  is involved in cell proliferation (GO:0008283). The function of this protein is not yet established \cite{li2013transcriptomic}, however NME2 has been linked to reduction of cell proliferation \cite{liu2015nme2} and proteins encoded in the neighbour locations of the DNA tend to have similar function \cite{feuerborn2015activity}. For microRNA mir--$3606$, we predict a role in collagen fibril organization (GO:0030199). Collagen plays a key role in cell adhesion, which can involve integrin \cite{jokinen2004integrin,testaz2001central} and mir--$3606$ has been linked to integrin in the literature as it has been suggested that mir--$3606$ can bind to ITGA4 (integrin subunit alpha 4) \cite{wong2014mirdb}. Finally, we propose that LOC101929876 (40S ribosomal protein S26) is involved in rRNA processing (GO:0006364), which is corroborated by the Reactome database in which the protein is associated with a major pathway of rRNA processing in the nucleolus and cytosol \cite{reactome1}.

These results confirm the ability of our hypergraphlets to predict biological functions of proteins from the wiring patterns in our novel model capturing multi-scale organisation of proteins in a cell.

\section{Conclusion}

We highlight the importance of considering the higher order organisation of protein interactions in conjunction with the standard PPI networks. We propose a novel methodology, hypergraphlets, to quantify the local wiring patterns of hypergraphs. We apply it to biological hypernetworks representing protein complexes and pathways of yeast and human and demonstrate a strong link between hypernetwork structure and the function of the proteins. Our novel methodology is able to mine the biological information hidden in the multi-scale architecture of molecular organisation. Furthermore, our analysis highlights the superiority, in terms of uncovering the underlying biology, of our multi-scale model when compared to the standard PPI networks. Additionally, we demonstrate that our new hypernetwork model, combined with our hypergraphlets, can be used for functional predictions. 
 
Despite a simple functional prediction approach, we obtain promising results when using hypergraphlets on our new multi-scale model for functional predictions. It would be interesting to train an advanced machine learning model, such as random forrest, using HDVs as features in an effort to improve predictions. Finally, we have demonstrated that the union of networks capturing the multi-scale molecular organisation is strongly linked to the underlying biology of the molecules. It would be interesting to further investigate if different data integration methods could lead to even more biologically relevant models.

\section*{Funding}
This work was supported by UCL Computer Science departmental funds, the European Research Council (ERC) Starting Independent Researcher Grant 278212, the European Research Council (ERC) Consolidator Grant 770827, the Serbian Ministry of Education and Science Project III44006, the Slovenian Research Agency project J1-8155 and the awards to establish the Farr Institute of Health Informatics Research, London, from the Medical Research Council, Arthritis Research UK, British Heart Foundation, Cancer Research UK, Chief Scientist Office, Economic and Social Research Council, Engineering and Physical Sciences Research Council, National Institute for Health Research, National Institute for Social Care and Health Research, and Wellcome Trust (grant MR/K006584/1) and UK Medical Research Council (MC\_U12266B).
  
\bibliographystyle{abbrv}

\bibliography{references}

\end{document}